\documentclass[useAMS,usenatbib]{mn2e}
\usepackage{amsmath,amstext,amsgen,amsbsy,amsopn,amsfonts,theorem}

\usepackage[pdftex]{graphicx}
\usepackage{xspace}
\usepackage{amsmath}
\usepackage{hyperref}
\usepackage{framed}
\usepackage{txfonts}
\usepackage{epstopdf}
\usepackage{gensymb}

\usepackage[normalem]{ulem}

\makeatletter
\def\mr@ignsp#1 {\ifx\:#1\@empty\else #1\expandafter\mr@ignsp\fi}%
\newcommand{\multiref}[1]{\begingroup
\xdef\mr@no@sparg{\expandafter\mr@ignsp#1 \: }%
\def\mr@comma{}%
\@for\mr@refs:=\mr@no@sparg\do{\mr@comma\def\mr@comma{,}\ref{\mr@refs}}%
\endgroup}
\makeatother

\newcommand{\hypref}[2]{\ifx\href\asklfhas #2\else\href{#1}{#2}\fi}
\newcommand{\Secref}[1]{Section~\multiref{#1}}
\newcommand{\secref}[1]{Sec.~\multiref{#1}}
\newcommand{\Appref}[1]{Appendix~\multiref{#1}}

\newcommand{\Tabref}[1]{Table~\multiref{#1}}

\newcommand{\Figref}[1]{Figure~\multiref{#1}}
\newcommand{\figref}[1]{Fig.~\multiref{#1}}
\renewcommand{\eqref}[1]{Equation (\multiref{#1})}

\newcommand{\eq}[1]{\begin{align}#1\end{align}}

\newcommand{\nln}{\nonumber\\}

\title[Number and size of gaps in stellar
  streams]{The number and size of
  subhalo-induced gaps in stellar streams}

\author[D. Erkal, V. Belokurov, J. Bovy, J. L. Sanders]
  {Denis Erkal$^1$\thanks{derkal@ast.cam.ac.uk}, Vasily Belokurov$^1$\thanks{vasily@ast.cam.ac.uk}, Jo Bovy$^2$\thanks{bovy@astro.utoronto.ca}, and Jason L. Sanders$^1$\thanks{jls@ast.cam.ac.uk} \\
  $^1$Institute of Astronomy, Madingley Road, Cambridge, CB3 0HA, UK \\ $^2$ Department of Astronomy and Astrophysics, University of Toronto, 50 St. George Street, Toronto, ON M5S 3H4, Canada}

\begin{document}

\label{firstpage}

\maketitle

\begin{abstract}
Ample observational capabilities exist today to detect the small
density perturbations that low-mass dark matter subhaloes impart on
stellar streams from disrupting Galactic satellites. In anticipation
of these observations, we investigate the expected number and size of
gaps by combining an analytic prescription for gap evolution on
circular orbits with the flux of subhaloes near the stream. We explore
the distribution of gap sizes and depths for a typical cold stream
around the Milky Way and find that for a given stream age and gap
depth, each subhalo mass produces a characteristic gap size. For a
stream with an age of a few Gyr, orbiting at a distance of 10-20 kpc
from the Galactic center, even modest subhaloes with a mass of
$10^6-10^7 M_\odot$ produce gaps with sizes that are on the order of
several degrees. We consider the number and distribution of gap sizes
created by subhaloes with masses $10^5-10^9 M_\odot$, accounting for the expected depletion of subhaloes by the Milky Way disk, and present
predictions for six cold streams around the Milky Way. For Pal 5, we
forecast 0.7 gaps with a density depletion of at least 25\% and a
typical gap size of $8^\circ$. Thus, there appears to be {\it no
  tension} between the recent non-detection of density depletions in
the Pal 5 tidal tails and $\Lambda$CDM expectations. These predictions
can be used to guide the scale of future gap searches.
\end{abstract}

\begin{keywords}
 galaxies: haloes - structure - cosmology: dark matter
\end{keywords}

\section{Introduction}

To date, only two promising techniques have been put forward to detect
{\it individual} low-mass dark matter (DM) clumps devoid of stars. The
presence of these so-called DM subhaloes may be betrayed by small
perturbations in the images of cosmological gravitational lenses
\citep[see e.g.][]{Mao1998, Dalal2002, Hezaveh2013}, or,
alternatively, be revealed by gaps in the stellar streams around the
Milky Way (MW) galaxy \citep[see
  e.g.][]{ibata_et_al_2002,johnston_et_al_2002,siegal_valluri_2008,carlberg_2009}. In
principle, through gravitational lensing it might be possible to
detect subhaloes with masses as low as $10^7 M_{\odot}$, and,
encouragingly, measurements have already been reported of DM subhaloes
with $M \lesssim 10^9 M_{\odot}$ \citep[see
  e.g.][]{Vegetti2010,Vegetti2012,Hezaveh2016}. This is reassuring, as
many dwarf galaxies have been shown to exist around the Milky Way with
masses similar to that or lower \citep[see
  e.g.][]{belokurov2013}. Moving forward, gaps in stellar streams
offer the possibility to pin down the DM mass spectrum below the dwarf
galaxy threshold, in other words in the completely dark regime
\citep[see e.g.][]{subhalo_properties}. The importance of such a
direct observational tool is difficult to overestimate, as
$\Lambda$CDM expects any MW-size galaxy to be bathed in a plethora of
DM subhaloes, with an overwhelming prevalence of low-mass objects
\citep[see e.g.][]{springel_et_al_2008}.

Over the last decade, detection of DM subhaloes via gaps in stellar
streams has evolved from a plausible idea into an imminent
measurement. This is because the haul of cold stellar streams (the
prime contender to carry marks of an interaction with low-mass DM
subhalos) has risen to at least a dozen structures
\citep{stream_book_grillmair_carlin} thanks to high-quality data from
all-sky imaging surveys like SDSS \citep[see e.g.][]{sdssdr9}, VST
ATLAS \citep[][]{shanks2015} and DES \citep[][]{des_survey}. While the
original survey data is clearly deep enough to identify the streams,
it is likely too shallow to warrant an unambiguous detection of
low-amplitude density fluctuations caused by DM subhalo
flybys. Nonetheless, candidate gaps have been reported in Pal 5 \citep{pal5disc} and
GD-1 \citep{gd1disc} streams using the SDSS photometry alone \citep[see
  e.g.][]{carlberg_pal5_2012,carlberg_gd1_2013}. Curiously, the
majority of these gaps are smaller than $2^{\circ}$ in size, seemingly
in agreement with $\Lambda$CDM-inspired predictions \citep[see
  e.g.][]{yoon_etal_2011,carlberg_2012,ngan2014}. The consensus in the
literature is that the cosmological structure-formation predictions
can be tested by measuring the shape and the normalisation of the gap
size spectrum. Thus, the intuition is that the incidence of gaps
encodes the sub-halo volume density in the vicinity of the stream, and
the gap size is linked to the mass of the dark perturber. Recently,
\cite{ibata_et_al_pal5} measured the stellar density along the Pal 5
stream to a significantly fainter magnitude limit. Interestingly, they
found no evidence for gaps on small scales.

It appears, therefore, that the preliminary studies of the density
fluctuations in the SDSS stellar streams might have erred on the side
of risk, when considering the possible false positives. This is
perhaps unsurprising as the interplay between the survey systematics
and the stellar halo density field at faint magnitudes has not been
studied in detail. The number of stars entering a particular
color-magnitude box (used to trace the stream) will depend on the
weather conditions at the epoch of observation. More precisely, sky
brightness and seeing will determine the object detectability and the
efficiency of star-galaxy classification. Unfortunately, only global
estimates of the SDSS completeness exist \citep[see
  e.g.][]{sdss_edr,sdss_dr2}. Temporal changes in completeness and
star-galaxy separation efficiency may therefore be reflected in
spatial variations of the faint star counts. For example, in Figure 1
of \citet{koposov_2012}, a stripy ``patchwork'' appearance of the SDSS
stellar density distribution can be observed, which remains visible -
albeit slightly subsided - even after application of the
``uber-calibration'' procedure to the SDSS photometry
\citep[][]{sdss_ubercal}. Coupled with weather conditions, SDSS survey
geometry can potentially induce spurious variations in the
density field of faint stars on a variety of angular scales. The SDSS
footprint consists of $2 \fdg 5$-wide stripes, each comprising of $1
\fdg 25$-wide strips. Thus, spurious power may be added on scales of
$2 \fdg 5$ degrees and down to a small fraction of a degree, the
latter due to the fact that individual stripes can overlap by
different amounts depending on the distance from the survey
poles. Worse still, because only a portion of a strip can be completed
during one night (an SDSS ``run''), bogus density fluctuations may
exist on scales of several to tens of degrees due to changes in epoch
(and hence weather conditions) along the individual
stripe. Furthermore, various other sources of spurious density
fluctuations are expected to exist, such as those associated with
large scale structure and saturated stars. Naturally, many authors
  attempt to mitigate against the above problems by limiting their
  stellar samples to brighter magnitudes, e.g. $r<22$. However, some
  of the issues discussed may unfortunately be exacerbated by the
  breakdown of the star-galaxy separation even at brighter
  magnitudes. While, globally, the SDSS completeness is 95$\%$ at
  $r=22.2$\footnote{Note, however, an increasingly erratic behaviour
    of the rms scatter in completeness estimate at $r=21$ in Figure 8
    of \citet{annis2014}}, the star-galaxy separation is 95$\%$
  correct at only $r=21$. This deteriorates to $90\%$ at $r=21.6$
\citep[see e.g.][]{annis2014}. This erroneous morphological
  classification can lead to spurious clumping in stellar density maps
  induced by the leakage of power from galaxy distributions as
  illustrated, for example, in Figure 4 of \cite{koposov2008}.

Deeper follow-up imaging with better seeing and under a darker sky,
additionally conforming to a different mosaic geometry, would naturally
do away with most of the artifacts discussed above. But if the number
and size of gaps reported earlier were in agreement with $\Lambda$CDM,
does it mean that the null detection reported by
\citet{ibata_et_al_pal5} is in tension with the predictions of
cosmological structure formation theory? Motivated by this conundrum,
we re-visit the expectations for the frequency and the scale of gap
creation. The first comprehensive attempt to describe the spectrum of
stream gaps due to interactions with DM subhaloes can be found in
\cite{yoon_etal_2011} who lay out a simple framework to count the
number of subhalo flybys near a stream. For example, for the Pal 5
stream, they estimated $\sim 5$ close flybys for subhaloes with masses
in the range of $10^7-10^8 M_\odot$. Exactly how large and deep a gap
these flybys would create, and hence how detectable they would be,
depends on the flyby geometry and the flyby
velocity. \cite{carlberg_2012} made the first attempt to answer this
question by combining a similar flyby counting technique as
\cite{yoon_etal_2011} with fits to the properties of gaps created by
subhalo flybys.

In this work, we will build on the approaches of \cite{yoon_etal_2011}
and \cite{carlberg_2012} by using a similar estimate for the number of
subhalo encounters, while determining the effect of each flyby based
on the results of \cite{three_phases}. Theirs is an analytic model of
the gap properties for density perturbations induced in streams on
circular orbit. The advantage of this approach is that the analytic
model works for any flyby geometry. This allows us to sample a wide
range of encounters and determine the distribution of gap properties
expected for a given stream and subhalo distribution. The assumptions
of this method, both the flyby rates and their properties, as well as
the gap properties, are tested against numerical simulations. Our
predictions also account for the expected depletion of subhaloes by
the Milky Way disk in the inner regions of the Milky Way
\citep{donghia_et_al_2010}. With this approach we find that
dramatically fewer gaps are expected than the results of
\cite{yoon_etal_2011} and \cite{carlberg_2012} suggested. In addition,
we present predictions for the distribution of gap sizes expected from
a $\Lambda$CDM spectrum of subhaloes and find that the characteristic
size is rarely lower than several degrees, i.e. typically, an order of
magnitude larger than that searched for by
\citet{ibata_et_al_pal5}. Thus this paper will demonstrate that the
lack of small-scale gaps in Pal 5 is not in tension with $\Lambda$CDM.

The paper is organized as follows. In \Secref{sec:flyby_properties}
the rate and properties of the subhalo flybys are derived. Next, in
\Secref{sec:method} we review and test the model of gap growth which
translates the flybys into gap properties. In
\Secref{sec:gap_properties} we use this formalism to examine how the
distribution of gap sizes and depths changes for various subhalo masses
and stream ages. We also give the distribution of gap sizes expected
from a $\Lambda$CDM population of subhaloes. In
\Secref{sec:gaps_in_real_streams} we predict the number of gaps in six
cold streams around the Milky Way and find significantly fewer gaps
than was previously expected. The model of the rate and properties of
the flybys, as well as the gaps they produce, is tested with N-body
simulations in \Secref{sec:testing}. We discuss implications of this
work for gap searches, possible contamination from giant molecular
clouds, and limitations of the method in \Secref{sec:discussion}. Finally, we conclude in
\Secref{sec:conclusion}.

\section{Setting up subhalo-stream encounters} \label{sec:flyby_properties}

In order to compute the expected number of stream gaps, we must first compute the expected number of subhalo flybys, as well as their velocity distribution relative to the stream. The expected number of flybys is controlled by the velocity distribution of the subhaloes as well as their number density. Our approach is similar to that in \cite{yoon_etal_2011} with several amendments. First, we will present a modified version of their derivation which correctly accounts for the velocity distribution of subhaloes. Second, we will use a lower number density of subhaloes since the presence of a baryonic disk in the Milky Way will deplete substructure by a factor of 2-3 \citep{donghia_et_al_2010}. Finally, we will use a higher subhalo velocity dispersion motivated by models of the Milky Way and cosmological simulations. Below, we will assume that the subhaloes are uniformly distributed and that each component of their velocity follows a normal distribution with a mean of zero and a dispersion of $\sigma$. This is sometimes referred to as an isotropic Maxwellian distribution. We will also neglect the size of the stream, treating it as a line, and assume that each star in the stream is just moving along this line, thus neglecting the velocity dispersion in the stream. 

\subsection{Expected number of flybys} \label{sec:num_flybys}

The effect of each subhalo flyby depends on the impact parameter to
the stream. Both \cite{yoon_etal_2011} and \cite{carlberg_2012} set up
a straightforward scheme to count the number of subhaloes which pass
within a given distance, $b_{\rm max}$, of the stream. We will now
present a slightly modified version of their calculation and explain
the difference with their result.

\begin{figure}
\centering
\includegraphics[width=0.5\textwidth]{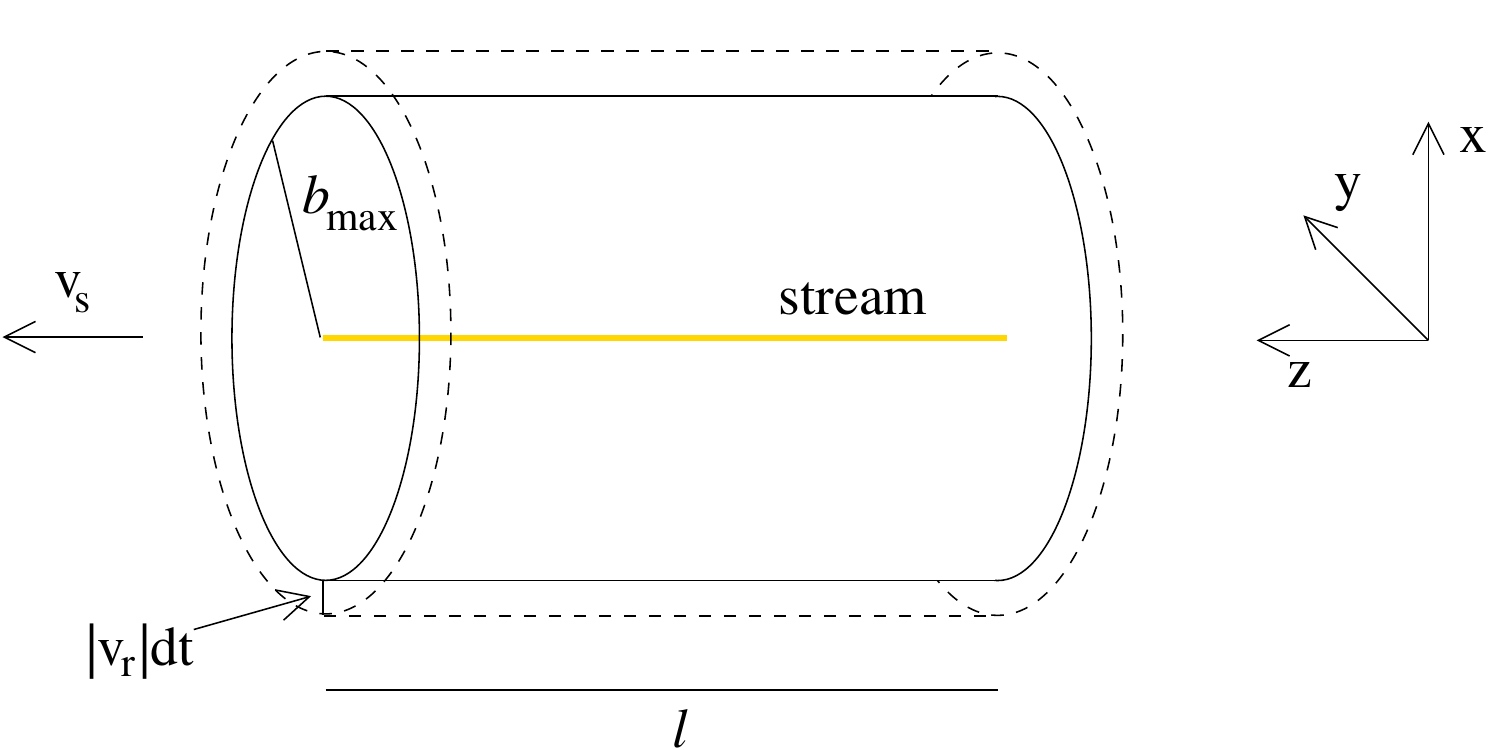}
\caption{Imaginary cylinder around stream used to count how many subhaloes pass near a stream. The yellow line represents the stream. The region between the solid cylinder and the dashed cylinder shows the volume from which subhaloes with radial velocity $v_r$ will enter within $b_{\rm max}$ of the stream in time $dt$. We restrict $v_r$ to be negative since we only want to count the subhaloes entering this cylinder and not those leaving. $v_s$ is the velocity of the stream.}
\label{fig:cylinder}
\end{figure}

We consider a cylinder of radius $b_{\rm max}$ around the stream, as shown in \Figref{fig:cylinder}. The number of subhaloes piercing this cylinder in some time interval, $dt$, is given by
\eq{ dN_{\rm enc} &= (2 \pi b_{\rm max} l) \times (| v_r | dt)\times n_{\rm sub}  \times P(v_r) dv_r , \label{eq:rate_Pvr} }
where $l$ is the length of the stream and $v_r$ is the cylindrical radial velocity in the stream coordinates, i.e. perpendicular to the motion of the stream. If we only consider the side of the cylinder and not the end caps, the flyby rate only depends on $v_r$. Since the radial velocity is just a projection of the cartesian velocities, $P(v_r)$ is also a Gaussian with a mean of zero and a dispersion of $\sigma$. 

Integrating over the negative radial velocities, i.e. those that are about to enter the cylinder, we get
\eq{ \frac{dN_{\rm enc}}{dt} = \sqrt{2\pi} \sigma b_{\rm max} l n_{\rm sub} . \label{eq:fiducial_rate} }
We can also compute the total number of encounters the stream will have by accounting for the growth of the stream in time. Assuming that the stream growth is linear in time, $l \propto t$, if we integrate the encounter rate until the present time, when the stream has a length of $l_{\rm obs}$, we get
\eq{ N_{\rm enc} = \sqrt{\frac{\pi}{2}} l_{\rm obs} b_{\rm max} n_{\rm sub} \sigma t  \label{eq:num_encounters}.}
A similar result is presented in \cite{yoon_etal_2011}, who used $l_{\rm obs}=4 (t/T_\psi) R_{\rm circ} \Delta \Psi$ where $\Delta \Psi$ is the angular growth per orbit, $T_\psi$ is the angular period, and $R_{\rm circ}$ is the orbital radius of the stream. Plugging this value of $l_{\rm obs}$ into \eqref{eq:num_encounters}, we get 
\eq{ N_{\rm enc} = 2\sqrt{2 \pi} R_{\rm circ} b_{\rm max} \sigma t n_{\rm sub} \Delta \Psi \Big( \frac{t}{T_\psi} \Big) .}
This can be now compared with equation 15 of \cite{yoon_etal_2011} in the limit that their encounter velocity, $v_{\rm enc}$, is taken to infinity, where we find that the number of encounters in this work is $2\sqrt{2}$ smaller. The difference is due to what is assumed about the radial velocity distribution, $P(v_r)$, which is used in \eqref{eq:rate_Pvr}. While we have argued that this distribution should be a Gaussian, \cite{yoon_etal_2011} instead used the relative speed distribution between two particles drawn from an isotropic Maxwellian distribution. This overestimates the radial velocity and hence the flux into the cylinder. We also note that our rate of flybys agrees with the rate per length derived in equation 3 of \cite{carlberg_2012}. 

\begin{figure}
\centering
\includegraphics[width=0.5\textwidth]{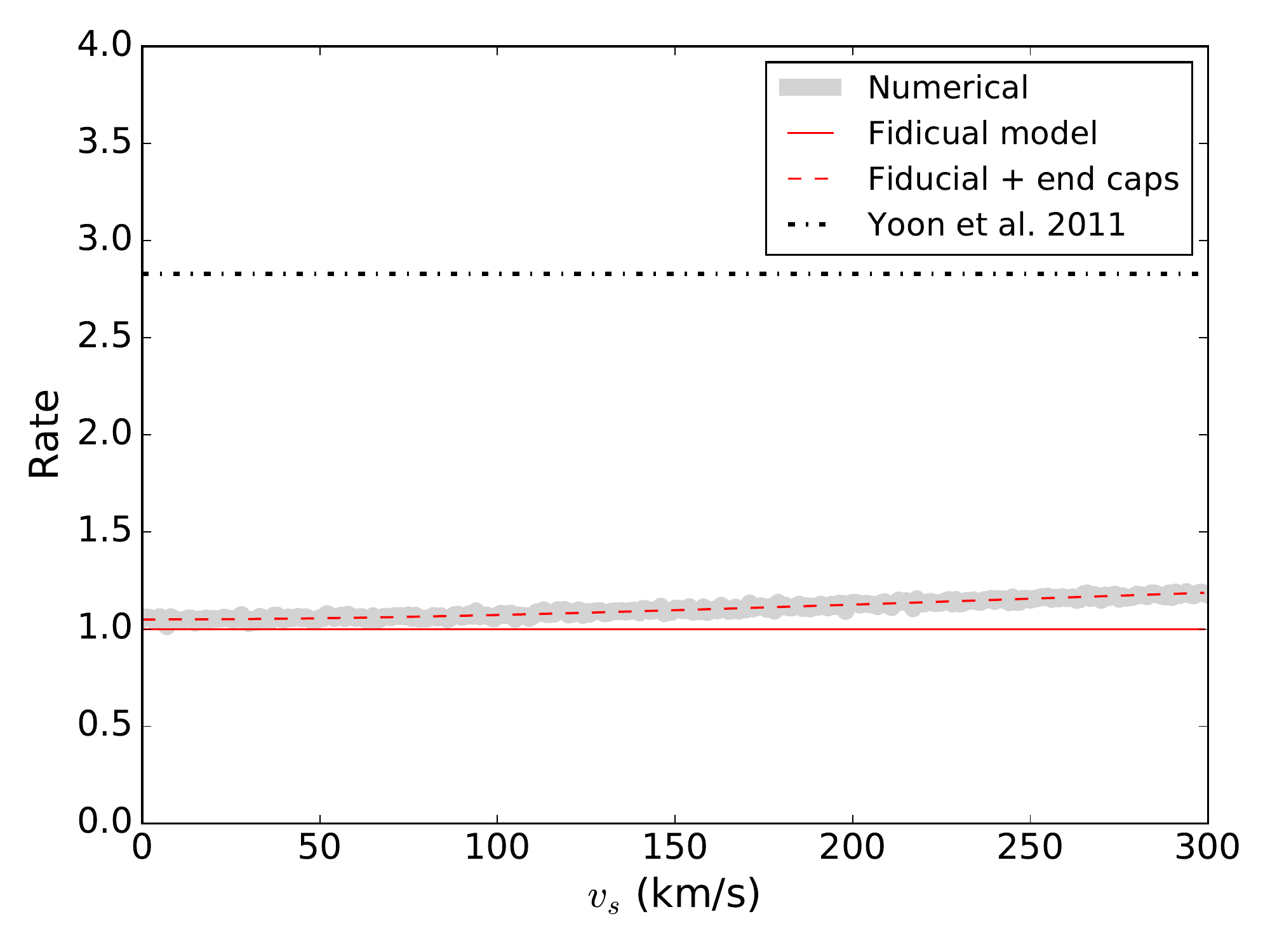}
\caption{Rate of particles entering a cylinder around the stream as a function of the stream's velocity through the cloud of particles. The light grey broad line shows the result of our simple numerical experiment. The red solid line shows our prediction from \protect\eqref{eq:num_encounters}. The red dashed line shows our fiducial model plus the contribution of subhaloes entering through the end caps of the cylinder which matches the numerical rate. We see that the fiducial model captures the bulk of the rate. Finally, the dot-dashed black curve is the rate from \protect\cite{yoon_etal_2011}. } 
\label{fig:numerical_flyby}
\end{figure}

To confirm that this rate is correct, we perform a simple numerical test. We take a cloud of particles with positions drawn from a uniform distribution. Each particle is assigned a velocity drawn from a normal distribution with $\sigma = 100$ km/s in each component. The particles are stepped forwards in time and we count the number of particles entering a cylinder, representing the region near a stream, which had a height of 20 and a radius of 1 in arbitrary units. We show the comparison in \Figref{fig:numerical_flyby} as a function of the stream velocity, $v_s$. We see that our simple model in \eqref{eq:num_encounters} captures the bulk of the numerical encounter rate. We also see that the numerical rate has a slight dependence on the stream velocity which is due to subhaloes passing through the end caps of the cylinder. In \Appref{sec:appendix} we derive the rate of subhaloes entering through the end caps which is presented in \eqref{eq:end_caps}. We show this model in \Figref{fig:numerical_flyby} as the dashed red line and we find that it matches the numerical result. 

\subsection{Velocity Distribution of Flybys}

\begin{figure}
\centering
\includegraphics[width=0.5\textwidth]{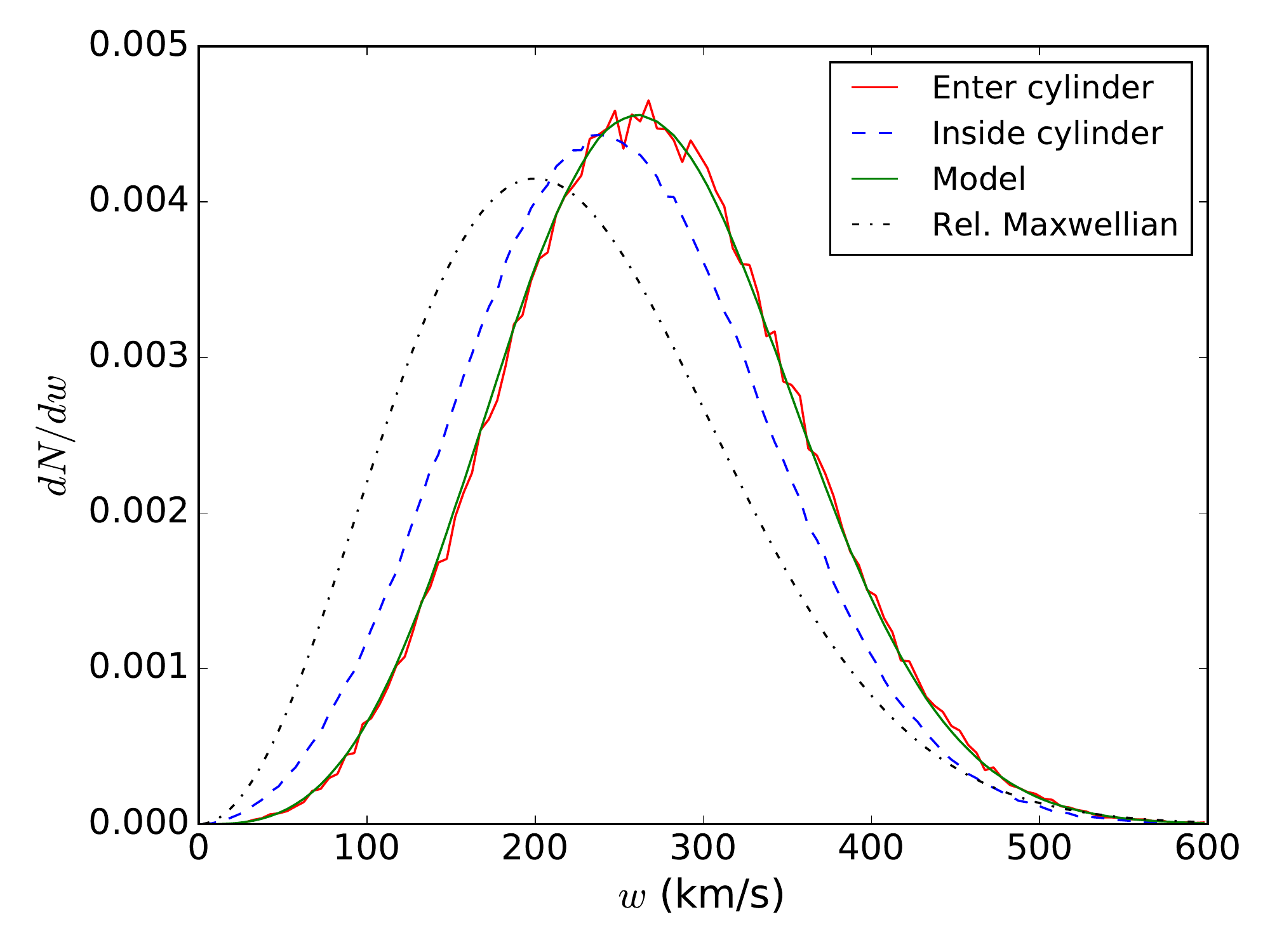}
\caption{Velocity distribution relative to stream in numerical example. The solid red curve shows the velocity distribution of particles which entered the cylinder in the time interval. The dashed blue curve shows the velocity distribution of the particles which were initially inside the cylinder, highlighting that it is different from the distribution of those which enter. The solid green curve shows the distribution of our model using \protect\eqref{eq:Ppara} and \protect\eqref{eq:Pperp}. The dot-dashed black curve shows the relative speed distribution of two particles drawn from an isotropic Maxwellian distribution used in \protect\cite{yoon_etal_2011}. In this example we have used a stream velocity of $v_s = 200$ km/s and a velocity dispersion of $\sigma = 100$ km/s.  } 
\label{fig:vdist_flyby}
\end{figure}

The distribution of flyby velocities is critical for understanding the gaps which subhaloes create since the encounter geometry and velocity affects the gap properties \citep{three_phases}. While the velocity distribution of subhaloes in the galaxy is given by a Gaussian in each direction, the distribution of the subhaloes that interact with the stream, i.e. those that pass within $b_{\rm max}$, is not. First, the stream is moving in a given direction. The relative velocity in this direction, $w_{\parallel}$, is given by
\eq{ P(w_\parallel|b_{\rm max}) = \frac{1}{\sqrt{2\pi \sigma^2}}  \exp\left(-\frac{(w_\parallel+v_s)^2}{2\sigma^2} \right) , \label{eq:Ppara}}
where $v_s$ is the velocity of the stream. The distribution of the velocity perpendicular to the stream's motion, $w_\perp$, can be derived by considering the distribution of the radial and tangential velocity relative to the stream. As subhaloes enter a cylinder with radius $b_{\rm max}$ around the stream, they have a radial velocity, $v_r$, and a tangential velocity, $v_\theta$. As we saw in \Secref{sec:num_flybys}, the rate at which particles enter this cylinder is proportional to their radial velocity, e.g. \eqref{eq:rate_Pvr}. Thus, the radial velocity distribution of subhaloes is given by
\eq{ P(v_r|b_{\rm max}) = \frac{|v_r|}{\sigma^2} \exp\left( -\frac{v_r^2}{2\sigma^2} \right) . \label{eq:Pvr}}
Note that this distribution is only nonzero for negative $v_r$, i.e. for the subhaloes heading towards the stream. The velocity distribution in the tangential direction, $P(v_\theta|b_{\rm max})$, is Gaussian with a mean of zero and a dispersion of $\sigma$. By combining these two velocities into $w_\perp = \sqrt{v_r^2 + v_\theta^2}$, we find
\eq{ P(w_\perp|b_{\rm max}) = \sqrt{\frac{2}{\pi}} \frac{w_\perp^2}{\sigma^3} \exp\left( - \frac{w_\perp^2}{2\sigma^2} \right) \label{eq:Pperp} . }
\eqref{eq:Ppara} and \eqref{eq:Pperp} give us the velocity distribution of the subhaloes which have passed near the stream. The dispersion of the flyby speed, $w=\sqrt{w_\parallel^2+w_\perp^2}$, is given by
\eq{ \langle w^2 \rangle = v_s^2 + 4 \sigma^2 .}
From \eqref{eq:Ppara} and \eqref{eq:Pperp} we see that the distribution of relative speeds is not simply the relative speed distribution of two particles drawn from an isotropic Maxwellian distribution. This is because the radial velocity distribution is biased since subhaloes with higher radial velocities towards the stream are more likely to enter a region near the stream, i.e. \eqref{eq:Pvr}. As in \Secref{sec:num_flybys}, we can check these velocity distributions against a numerical example of subhaloes distributed uniformly in position with an isotropic Maxwellian velocity distribution and look at the properties of subhaloes which enter a cylinder around the stream. In \Figref{fig:vdist_flyby} we compare the numerically derived velocity distribution against our model and find excellent agreement. For contrast, we also show the velocity distribution of particles inside the cylinder and the relative speed distribution assumed in \cite{yoon_etal_2011}. We also show the velocity distribution of the particles which were initially in the cylinder to emphasize that it is different from those entering the cylinder. For this example we used a velocity dispersion of $\sigma = 100$ km/s, a cylinder with an aspect ratio of 1:10, and a stream velocity of $v_s = 200$ km/s. 

\subsection{Impact parameter distribution}

Now that we have the velocity distribution of particles which enter a cylinder with radius $b_{\rm max}$ around the stream, we can compute the distribution of impact parameters. A particle which enters the cylinder with a radial velocity of $v_r$ and a tangential velocity of $v_\theta$ will have an impact parameter of 
\eq{ b = b_{\rm max} \frac{ | v_\theta |}{\sqrt{v_\theta^2+v_r^2} } . }
We can then determine $P(b|b_{\rm max})$, the distribution of the
impact parameters for the flybys within $b_{\rm max}$, by integrating
the velocity distributions over all $v_r, v_\theta$ which have an
impact parameter of $b$, i.e.
\eq{ P(b|b_{\rm max}) &= \int  P(v_r|b_{\rm max}) P(v_\theta|b_{\rm max}) \delta(b - b_{\rm max} \frac{ |v_\theta |}{\sqrt{v_\theta^2+v_r^2} } ) dv_r dv_\theta , \nln &= \frac{1}{b_{\rm max}}  . }
Thus we see that the impact parameters are distributed uniformly from $0$ to $b_{\rm max}$. At first sight, this result may seem counterintuitive since if we looked at the subhaloes near the stream at any particular time, the distribution of their distances would increase linearly with distance from the stream. However, the impact parameter is the minimum distance between the subhalo's path and the stream track. As such, the impact parameter distribution is really the distribution of distances between two lines in three dimensions, which is independent of distance. Both \cite{yoon_etal_2011} and \cite{carlberg_2012} also used a uniform distribution of impact parameters.

\subsection{Number density of subhaloes} \label{sec:num_density_subhaloes}

Next, we need an estimate of the number density of subhaloes. \cite{springel_et_al_2008} studied the number density profile of subhaloes around a Milky Way-like analogue and found that it is well described by an Einasto profile:
\eq{ n_{\rm sub} \propto \exp\bigg( - \frac{2}{\alpha} \Big( (\frac{r}{r_{-2}})^\alpha - 1 \Big) \bigg) , \label{eq:einasto}}
with $\alpha = 0.678$ and $r_{-2} = 0.81 r_{\rm 200} = 199$ kpc. They found that this fit worked for all mass ranges of subhaloes they explored, i.e. those from $10^5 M_\odot$ to $10^{10} M_\odot$. The host halo they simulated had a mass of $M_{\rm 200} = 1.839 \times 10^{12} M_\odot$. We will scale this down to a mass of $M_{200} = 10^{12} M_\odot$, reducing $r_{200}$ by $M^\frac{1}{3}$ and assume that the scaled down halo has the same concentration and same $\alpha$. Thus we would expect the same fit with $r_{-2} = 162.4$ kpc. Next we have the spectrum and normalization of the subhaloes. As in \cite{springel_et_al_2008}, we express the subhalo mass function as 
\eq{ \frac{dN_{\rm sub}}{dM} = a_0 \Big( \frac{M}{m_0} \Big)^n , }
with $a_0 = 3.26 \times 10^{-5} M_\odot^{-1}$, $m_0 = 2.52 \times 10^7 M_\odot$, and $n=-1.9$. Note that this was for the total number within $r_{\rm 50} = 433$ kpc. Scaling down to a MW mass of $10^{12} M_\odot$ (i.e. scaling the virial radius by $M^{1/3}$ and the number of subhaloes within the virial radius by $M$), we would get $a_0 = 1.77 \times 10^{-5} M_\odot^{-1}$ within $353$ kpc. This can now be combined with the density profile to get the correctly normalized subhalo profile
\eq{ \frac{dn_{\rm sub}}{dM} = c_0 \Big( \frac{M}{m_0} \Big)^n \exp\bigg( - \frac{2}{\alpha} \Big( (\frac{r}{r_{-2}})^\alpha - 1 \Big) \bigg) \label{eq:nsubM} ,}
with $c_0 = 2.02\times 10^{-13} M_\odot^{-1} {\rm kpc}^{-3}$. Using \eqref{eq:nsubM}, the number density of subhaloes in any mass range and location can be computed. We note that we found broadly similar results in the public catalogues of Via Lactea II \citep[VLII][]{Diemand2008} when we looked at the number of subhaloes within 50 kpc.

This estimate of the number density of subhaloes from the Aquarius simulations \citep{springel_et_al_2008} is based on a collisionless $N$-body simulation which neglects baryonic effects. \cite{donghia_et_al_2010} found that the presence of a disk with a mass of 10\% of that of the host galaxy decreases the number of subhaloes at $10^7 M_\odot$ by a factor of 3. Similar reductions are found over a wide range of masses so we will assume that the disk decreases the abundance of all subhaloes in the inner region by a factor of 3. Furthermore, we assume that the presence of the disk only changes the normalization of the number density of subhaloes but not its shape. As such, we account for the disk's presence by simply dividing the right-hand side of \eqref{eq:nsubM} by 3.

For the properties of individual subhaloes, we make fits to the $M_{\rm tidal}$-$v_{\rm max}$ relation to the subhaloes in VLII \citep{Diemand2008}. If we model the subhaloes as Plummer spheres, where we take the Plummer sphere mass to be $M_{\rm tidal}$, this gives a scale radius of 
\eq{ r_s = 1.62 {\rm kpc}~\Big( \frac{M_{\rm sub}}{10^8 M_\odot} \Big)^{0.5} . \label{eq:plummer_r} }
If the fit is instead made for a Hernquist profile, the relation would be
\eq{ r_s = 1.05 {\rm kpc}~\Big( \frac{M_{\rm sub}}{10^8 M_\odot} \Big)^{0.5} . \label{eq:hernquist_r}}

\subsection{Velocity distribution of subhaloes}

In addition to the number density, we must also specify the velocity distribution of subhaloes. As described above, we assume that the velocity distribution of each component is a Gaussian with a mean of zero and a dispersion of $\sigma$. This simplification neglects the velocity anisotropy of subhaloes seen in simulations, as well as the fact that the velocity distributions in simulations are not Gaussian \citep[e.g.][]{diemand_et_al_2004}. In order to make a prediction for the gap properties in streams around our Galaxy, we need an estimate of this dispersion for the inner region, $r<30$ kpc, of the Milky Way. 

Observationally, the radial velocity dispersion has been measured for a collection of stars, globular clusters, and satellite galaxies and a value of $\sigma \sim 120$ km/s within $30$ kpc was found \citep{battaglia_et_al_2005}. This value was also used as the fiducial subhalo velocity dispersion by both \cite{yoon_etal_2011} and \cite{carlberg_2012}. However, the velocity dispersion of subhaloes appears to be substantially higher than this. \cite{piffl_et_al_2015} constructed self-consistent equilibrium models for the Milky Way and found that the dark matter had velocity dispersions of 150-205 km/s near the location of the Sun. \cite{diemand_et_al_2004} compared the velocity dispersions of dark matter particles and subhaloes in cosmological simulations and found that the subhaloes had a velocity dispersion which is $\sim 10\%$ higher. The velocity dispersion can also be computed from cosmological simulations of Milky Way-like galaxies. Using the public catalogues of VLII \citep{Diemand2008}, we find velocity dispersions of 160-200 km/s in the three cartesian velocity components for subhaloes within 30 kpc of the Milky Way-analogue, although we note that those simulations do not include the effect of the disk and the halo is more massive than the Milky Way. With these results in mind we take $\sigma = 180$ km/s as our velocity dispersion for each velocity component. It is not immediately clear what this increased velocity dispersion means for the number of gaps since while it will result in a larger number of flybys, i.e. \eqref{eq:num_encounters}, it will also increase the relative speed of the flybys, i.e. \eqref{eq:Ppara,eq:Pperp}, which results in smaller perturbations to the stream and less pronounced gaps. However, the effect of the velocity dispersion was investigated in \cite{bovy_erkal_sanders} where they found that the main effect is from the number of flybys and the change in the gap properties is subdominant. 

\subsection{Evolution of the number density of subhaloes}

Streams are sensitive to the number density of subhaloes they
encounter starting from the epoch of the onset of the progenitor's
disruption to present day. However, they only interact with subhaloes
in the radial range which the stream
explores. \cite{diemand_et_al_2007} studied the number of subhaloes
within a fixed mass aperture in a Milky Way analogue. Their Figure 6
demonstrates that the number of subhaloes within a shell containing
the mass fraction $M/M_{200} < 1/6$ decreases by a factor of almost 3
from $z=1$ to $z=0$. They show that this mass shell has stabilized
between $z=3$ and $z=2$ so this estimate of the subhalo disruption can
also be thought of as for a given radial range. Many of the streams in
the Milky Way have been disrupting for a period of time similar to
this, and, thus, the change of the subhalo number density with
redshift should be taken in account. Although not shown in this work,
our numerical experiments indicated that the effect on a stream like
Pal 5 is not very significant, likely due to the relatively young age
of 3.4 Gyr we assume \citep[motivated by the results of][]{kuepper_et_al_pal5}, and thus currently we choose to ignore
it. However the older streams like GD-1 might be more affected. We
will come back to the importance of the subhalo number density
evolution with redshift in future work.

\section{Stream Gap Fabrication} \label{sec:method}

Now that we have computed the rate of flybys, we need a model for the effect of each encounter to determine the detectability of the gap it produces, and hence the number of gaps expected. \cite{three_phases} presented an analytic model for the evolution of a stream gap after a flyby with an arbitrary geometry, and provided analytic expressions for the width and depth of the gap. As in that work, we define the gap depth as $f\equiv \rho/\rho_0$ where $\rho$ and $\rho_0$ are the minimum perturbed density and the unperturbed density respectively. By combining this analytic model with the rate and properties of the flybys described in the previous Section, we can make a prediction for the properties of the gaps. We will provide a forecast for the expected number of gaps deeper than some density threshold, $f < f_{\rm cut}$ . A similar approach was taken in \cite{carlberg_2012} where a rate calculation was combined with fits to the gaps created by N-body simulations of flybys. We note that the gap widths and depths we consider are for an observer at the center of the Galaxy.

\subsection{Review of gap evolution}

\cite{three_phases} found that the gap growth proceeds in three phases. The first phase, the compression stage, is short-lived and leads to a minor increase in the density near the point of closest approach. We ignore this short-lived phase and instead focus on the second and third phases, the expansion and the caustic phase respectively, where the gap is created and then becomes wider and deeper. 

During both the expansion and the caustic phase, the gap depth, is given by
\eq{ \frac{\rho}{\rho_0} = \left( 1 + \frac{4-\gamma^2}{\gamma^2} \frac{w_\perp^2}{w^3} \frac{2 G M}{b^2+r_s^2} t \right)^{-1} , \label{eq:gap_density}}
where $\gamma$ is related to the host's gravitational potential, $\psi(r)$:
\eq{ \gamma^2 = 3 + \frac{r \partial_r^2 \psi(r)}{ \partial_r \psi(r) } . \label{eq:gamma}}
During the expansion phase, the size of the gap is given by
\eq{ \Delta \psi_{\rm gap} = 2 \frac{w}{w_\perp} \frac{\sqrt{r_s^2+b^2}}{r_0} + \frac{2 G M w_\perp}{w^2 r_0 \sqrt{r_s^2+b^2} }\frac{4-\gamma^2}{\gamma^2} t . \label{eq:gap_size_grow} }
The expansion phase continues until the caustic timescale, 
\eq{ t_{\rm caustic} = \frac{4 \gamma^2}{4-\gamma^2} \frac{w^3}{w_\perp^2} \frac{b^2+r_s^2}{GM} , }
after which the caustic phase begins and the gap size is given by
\eq{ \Delta \psi_{\rm gap} = 4 \left( \frac{4-\gamma^2}{\gamma^2} \frac{2 G M}{w r_0^2} t \right)^{\frac{1}{2}} . \label{eq:gap_size_caustic}}
These expressions describe the average time evolution of the gap density and size. Thus, given any impact, we can rapidly compute the density and size of the gap. By integrating over the impact parameter, subhalo velocity, and impact time, we can determine the distribution of gaps that are created. In this work we will assume that the rotation curve is locally flat and hence that $\gamma^2 = 2$. 

These expressions were derived assuming that the unperturbed stream is on a circular orbit and neglect both the eccentricity of the stream's orbit, and the energy and angular momentum dispersion in the stream. These effects were studied in \cite{sanders_bovy_erkal_2015} which found that while the picture in \cite{three_phases} is mostly correct, the dispersion in the stream can cause the density in gaps to plateau and that gaps can grow at slightly different rates depending on their location along the stream due to energy sorting of debris.

\subsection{Effective N-body Simulation} \label{sec:effective_nbody_vel_kicks}

In order to test the limits of the analytic picture above, we need to compare it against simulations. Since an N-body disruption of a globular cluster progenitor can take several tens of CPU hours to run, it is not feasible to investigate a large number of flybys. Instead, we have developed an effective N-body simulation where we first run a disruption of a progenitor to produce a stream. We then take an earlier snapshot of the simulation and compute the velocity kicks from a single subhalo flyby using the impulse approximation, accounting for the stream curvature as in \cite{sanders_bovy_erkal_2015}. The particles whose fractional energy change due to the kick is larger than some threshold are then evolved to the final time as tracers in the host potential. The particles within the progenitor do not receive a kick since they will be affected by the gravitational field of the progenitor and cannot be treated as tracers. At the final time, the perturbed particles are combined with the unperturbed particles from the final snapshot to give all particles in the stream. This method allows us to rapidly evaluate the gap profile from a variety of impacts.

For the N-body simulation, we simulate a Pal 5-like stream whose progenitor matches the measured line-of-sight velocity and proper motions in \cite{kuepper_et_al_pal5} at the present time. These simulations are performed with the N-body part of \textsc{gadget-3} which is similar to \textsc{gadget-2} \citep{springel_2005}. The progenitor is modelled as a King profile with a mass of $2\times 10^4 M_\odot$, a scale radius of $15$ pc, $w=2$, and is modelled with $10^5$ equal mass particles and a softening of $1$ pc. The progenitor is evolved in the \texttt{MWPotential2014} potential given in \cite{bovy_galpy} with the bulge replaced with a Hernquist profile with $M=5 \times 10^9 M_\odot$ and a scale radius of $0.5$ kpc. The simulation is run for 5 Gyr and snapshots are stored every 10 Myr.

\begin{figure}
\centering
\includegraphics[width=0.5\textwidth]{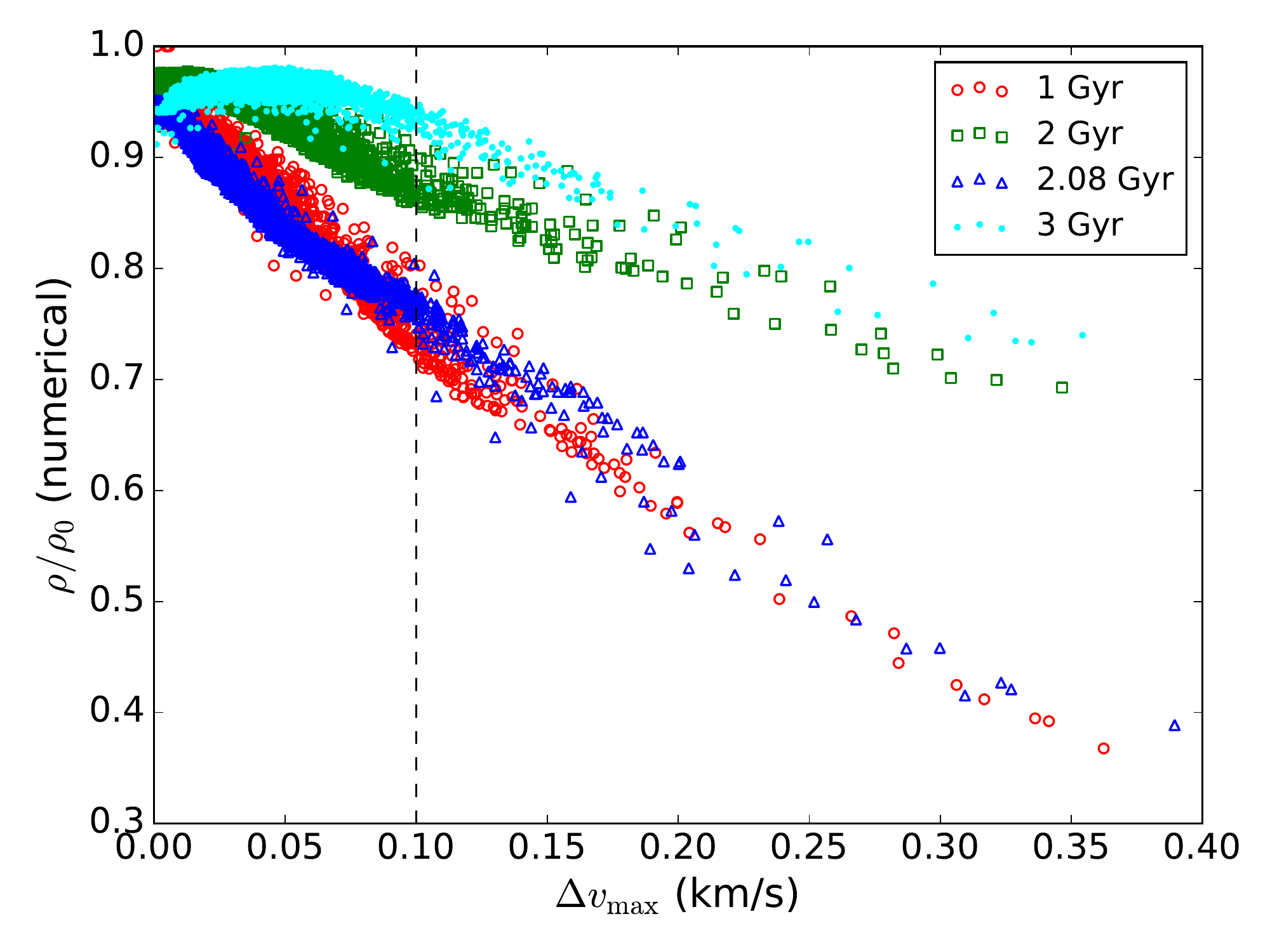}
\caption{Comparison of density from effective N-body simulations with maximum velocity kick for flybys of subhaloes with a mass of $1.6 \times 10^6 M_\odot$ evolved for varying amounts of time. We see that low velocity kicks do not produce a substantial gap and as the velocity increases, the gaps get deeper. We chose 0.1 km/s (the vertical black dashed line) as the cutoff since for these velocities we reliably get a significant depletion. The velocity kick where appreciable gaps are produced is similar for a range of masses so we use the same cutoff for all masses. The large difference between the gaps produced 2 Gyr ago and 2.08 Gyr is due to the stream being at a different orbital phase. } 
\label{fig:1e6flyby_deltav}
\end{figure}

We considered a wide range of impactors from $10^5$ to $10^8 M_\odot$ and found that evolving the particles whose change in energy exceeded a threshold of $\Delta E/E \geq 0.1$ was sufficient to reproduce the effect of the flyby. These simulations allow us to gauge when our simple analytic model begins to break down, i.e. when the velocity kicks become sufficiently small that the velocity dispersion in the stream becomes important, as well as when particles begin to fill in the gap. Specifically, we consider a $1.6 \times 10^6 M_\odot$ subhalo and sample a grid of subhalo velocities in each direction, $-500$ to $400$ km/s in steps of $100$ km/s, a grid of impact parameters, from 0 to 1 kpc in steps of $0.1$ kpc, and four different impact times of 1,2,2.08, and 3 Gyr ago. We find that the flybys which have a maximum velocity kick of $\Delta v \sim 0.1$ km/s can still produce an appreciable gap depth as shown in \Figref{fig:1e6flyby_deltav} where we compare the maximum velocity kicks for gaps of different ages. Somewhat surprisingly, this is substantially smaller than the velocity dispersion in the stream, $\sigma \sim 1$ km/s. We note that this cutoff to produce appreciable gaps will depend on the mass of the progenitor.

\begin{figure}
\centering
\includegraphics[width=0.5\textwidth]{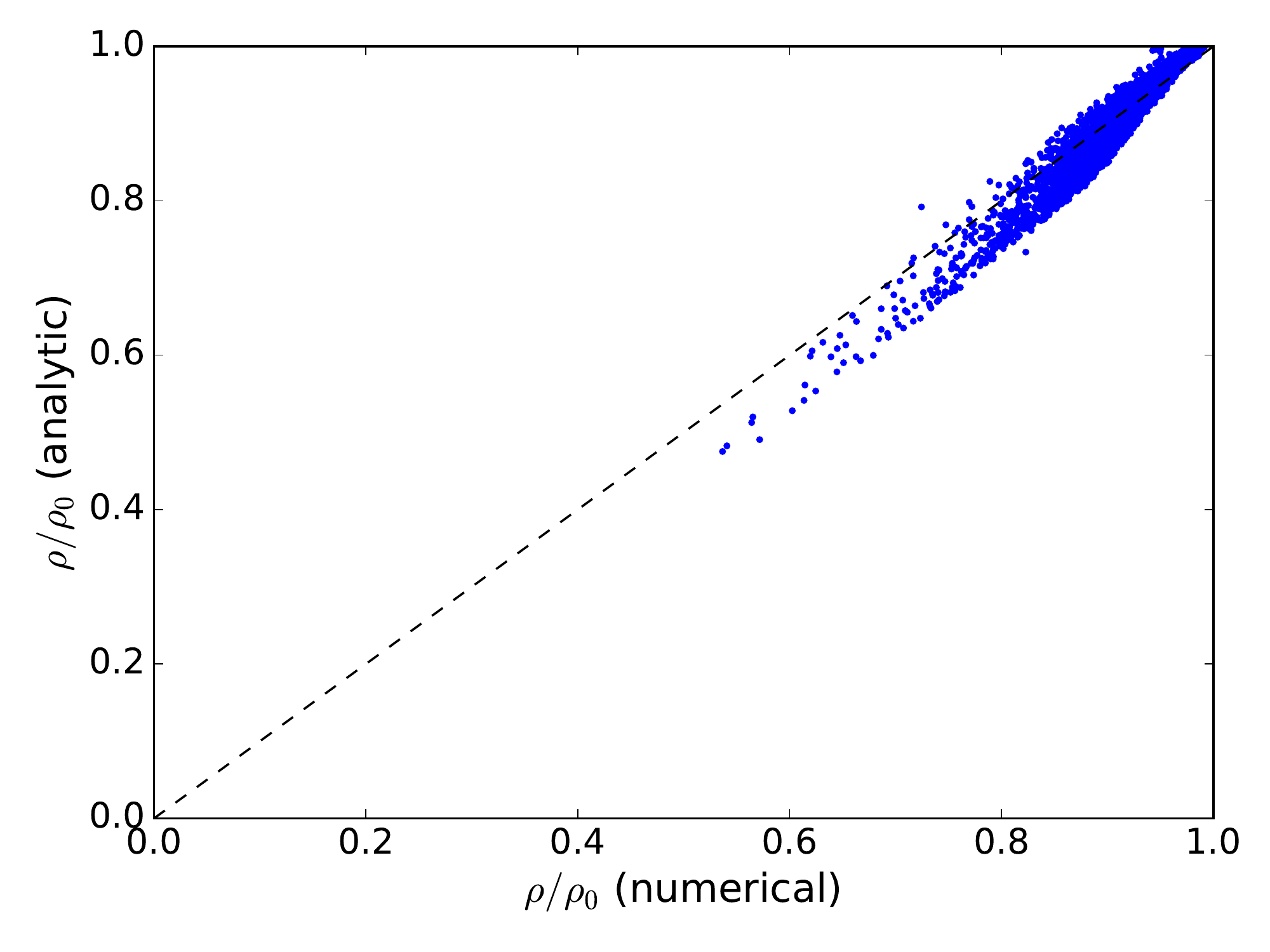}
\caption{Comparison of analytic density result versus the densities computed with the effective N-body for subhaloes with a mass of $10^6 M_\odot$ 1 Gyr after impact. The analytic result matches the numerically computed density, justifying the method. The match is similar for more massive subhaloes.} 
\label{fig:1e6flyby}
\end{figure}

The same simulations can be used to test the density formula given in \eqref{eq:gap_density}. We use the same grid as described above for the impacts 1 Gyr ago. For each flyby, the expected density using \eqref{eq:gap_density} is also computed. These are compared in \Figref{fig:1e6flyby} which demonstrates that while the gaps do get filled in (the numerical density is slightly higher than the analytic approximation), the agreement is rather good. We note that the level of agreement depends on the phase at which the gap is observed and that the numerical gap depth oscillates between being deeper and shallower than our analytic prediction depending on this phase. Thus our analytic model should be thought of as giving the average evolution of the gap depth. 

\section{Expected Properties of stream gaps} \label{sec:gap_properties}

With the method now in place, the properties of gaps created by a distribution of subhaloes can be explored. First, we consider a single subhalo and find the properties of the gaps it creates. It turns out that the stream gaps produced by a single DM clump have a characteristic size which depends on the age of the stream. Then, we consider subhaloes with masses drawn from the $\Lambda$CDM distribution and examine the properties of the gaps thus created. In this section, the properties of the fiducial stream model, i.e. its age, velocity dispersion, and orbital properties, are chosen to be similar to those of the Pal 5 stream given in \cite{kuepper_et_al_pal5}, namely we assume the age of the stream is 3.4 Gyr and that it is on a circular orbit with a radius of 13 kpc and a velocity of 220 km/s \citep{bovy_et_al_2012}.

\subsection{Gap density threshold and minimum gap size} 

When discussing observable gaps, we must introduce a density threshold below which the gap can be detected. As we saw in \Secref{sec:method}, subhalo flybys can produce arbitrarily shallow gaps which evidently will not be observable. However, since the gap size and gap depth both grow with time, in order for a gap to be deeper than some threshold, the gap must have grown to a certain extent. Thus, gap density and gap width are closely linked, and a gap density threshold gives a corresponding minimum gap size. Expressions for the gap size are given in \eqref{eq:gap_size_grow} and \eqref{eq:gap_size_caustic}. If the gap has a depth of $f=\rho/\rho_0$, this gives a gap size of 
\eq{ \Delta \psi_{\rm gap} = B\left(1 + f^{-1}\right) , \label{eq:gap_size_thresh_grow}}
in the expansion phase and 
\eq{ \Delta \psi_{\rm gap} = 4 B \sqrt{f^{-1}-1} ,  \label{eq:gap_size_thresh_caustic} }
in the caustic phase, where
\eq{ B = \frac{\sqrt{b^2+r_s^2}}{r_0} \frac{w}{w_\perp} .} 
Thus we see that a deep gap with $f \sim 0.5$ will have a gap size of several $B$ as seen from the center of the galaxy. The value of $B$ is the smallest for a direct impact where the relative velocity is perpendicular to the stream's motion. In this case, $B = \frac{r_s}{r_0}$. For example, if the stream is located at a typical Galactocentric distance we assume for Pal 5, $r_0 \sim 13$ kpc, a modest impactor with a mass of $10^7 M_\odot$ with a scale radius of $r_s \sim 500$ pc will have $B = 2.2^\circ$. Likewise, a smaller subhalo with a mass of $10^6 M_\odot$ and a scale radius of $r_s = 100$ pc will have $B = 0.4^\circ$. Therefore, it is clear that the smallest gap size will be at least $1^\circ$ and likely larger since most of the impacts will not be direct and will have velocity components which are aligned with the stream's motion. As a result, it is required that deep gaps should be fairly large, at least several degrees, or, in other words, deep gaps smaller than this size are not expected. 

\begin{figure}
\centering
\includegraphics[width=0.5\textwidth]{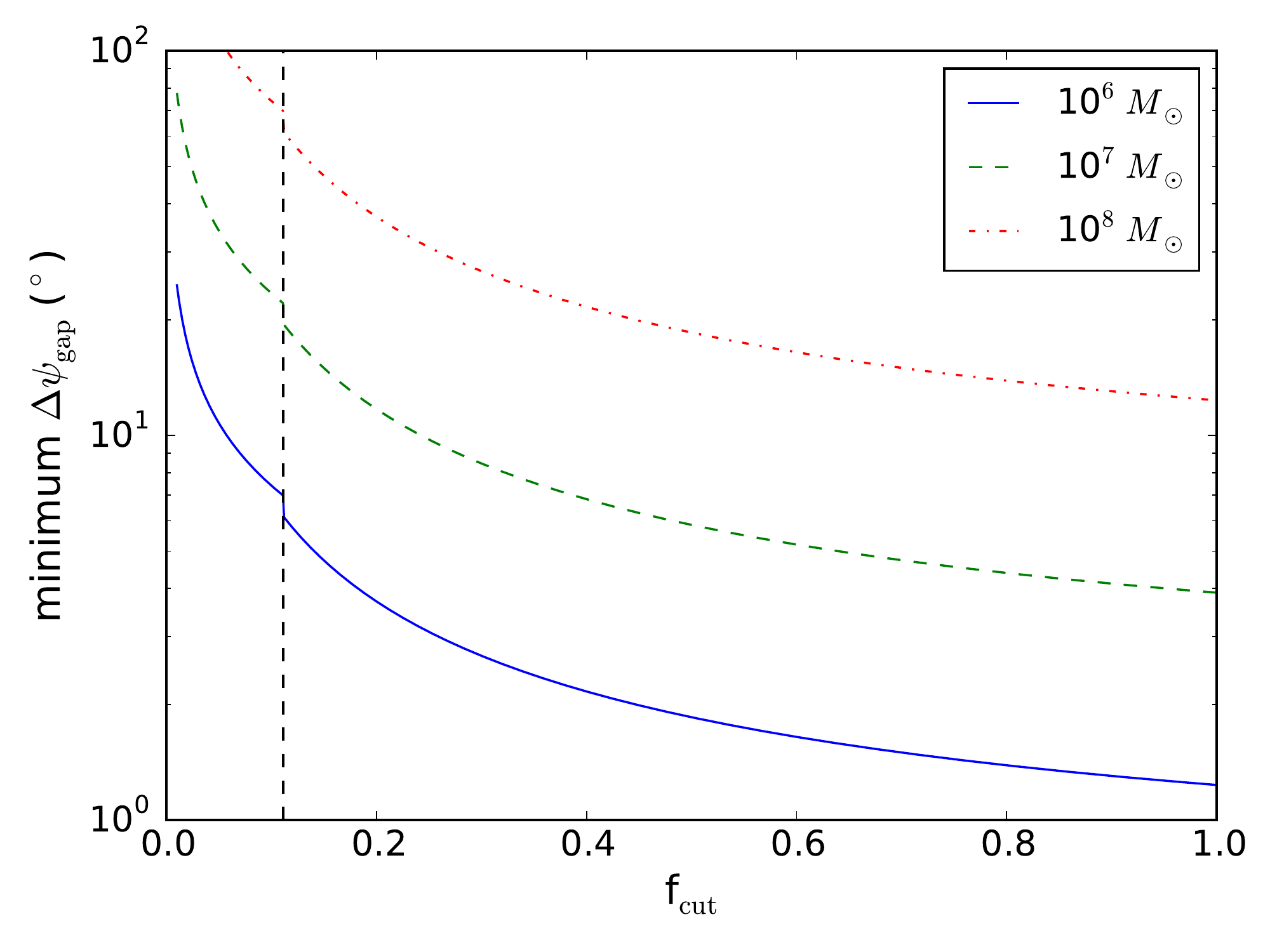}
\caption{Minimum gap size as a function of density in gap center for various mass subhaloes. These curves come from \protect\eqref{eq:gap_size_thresh_grow} for the growth phase with $f_{\rm cut} > 1/9$ and \protect\eqref{eq:gap_size_thresh_caustic} for the caustic phase with $f_{\rm cut} < 1/9$. The discontinuity at $f_{\rm cut} = 1/9$ is due to the fact that the leading order contribution to the gap size is not continuous as the gap progresses from the expansion to the caustic phase.} 
\label{fig:minimum_gap_size}
\end{figure}

\Figref{fig:minimum_gap_size} shows how the minimum gap size depends on the gap density threshold, $f_{\rm cut}$. As evidenced by the Figure, even for a $10^6 M_\odot$ subhalo, the minimum gap size is larger than $1^\circ$, while an object with a mass of $10^7 M_\odot$ will produce gaps in excess of $4^\circ$.

\subsection{Gap size distribution for single subhalo population} \label{sec:gap_size_dist_homogenous}

Let us now explore the gap size distribution created by a homogenous population of subhaloes with a single mass and scale radius. This can be achieved by sampling the distribution of flyby velocities and impact parameters. In practice, we randomly draw the parallel and perpendicular flyby velocities from 
\eqref{eq:Ppara} and \eqref{eq:Pperp} respectively, and the impact parameter from a uniform distribution between $0$ and $b_{\rm max}=5 r_s$. Finally, we draw the impact epoch from a linear distribution since the stream grows roughly linearly in time \citep[e.g. Fig. 2 of][]{bovy_erkal_sanders}. 

\begin{figure}
\centering
\includegraphics[width=0.5\textwidth]{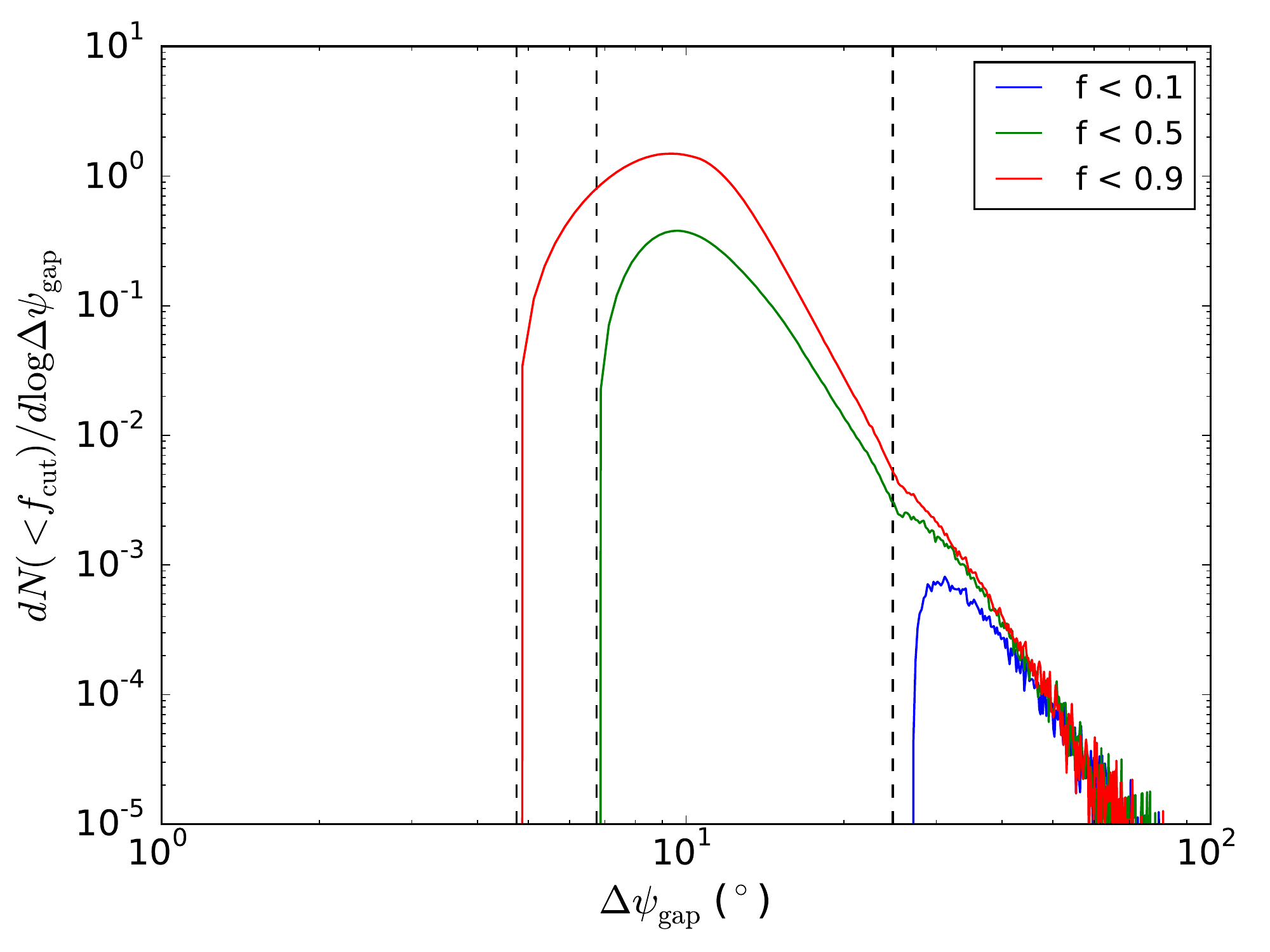}
\caption{Distribution of gap sizes for a subhalo with $M=10^7 M_\odot$ and $r_s = 512$ pc with varying gap density thresholds. The vertical dashed black lines come from \protect\eqref{eq:gap_size_thresh_grow}. We see that as the threshold decreases, the minimum gap size increases. This is expected since gaps which have had time to grow sufficiently deep will also have grown sufficiently large.  } 
\label{fig:gap_size_dist}
\end{figure}

\Figref{fig:gap_size_dist} shows the gap size distribution for three different density thresholds, $f_c$. The dashed vertical lines show the minimum gap size given by \eqref{eq:gap_size_thresh_grow}. We use a Plummer sphere subhalo with $M=10^7 M_\odot$ and a size of $r_s = 512$ pc. We see that as we decrease the density threshold, i.e. as we require deeper gaps, the gap size increases since in the time it takes the gap to achieve such a depth, it will also have grown to a substantial size. 

\begin{figure}
\centering
\includegraphics[width=0.5\textwidth]{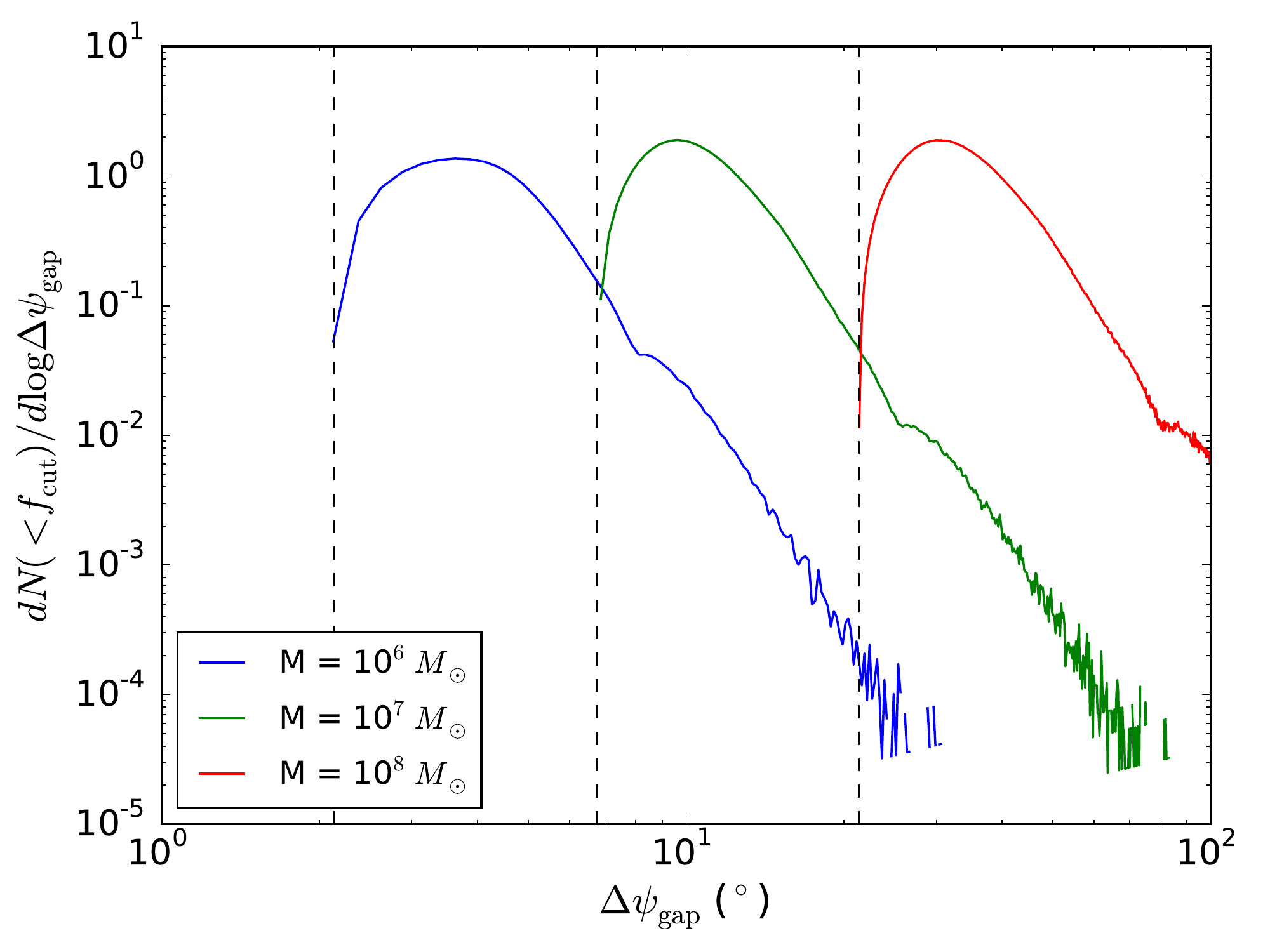}
\caption{Distribution of gap sizes created by subhaloes with varying mass and scale radii which create a gap deeper than 50\%. The vertical dashed black lines come from \protect\eqref{eq:gap_size_thresh_grow}. We see that as we increase the mass of the subhalo, we get a corresponding increase in the gap size. We also see that each mass subhalo creates gaps of a characteristic size. We emphasize that these gaps are quite large: even the $10^6 M_\odot$ subhalo creates gaps with a characteristic size of $\sim 3^\circ$.} 
\label{fig:gap_size_dist_vary_M}
\end{figure}

In \Figref{fig:gap_size_dist_vary_M} we examine how the gap size varies as we change the subhalo properties. We consider three different subhaloes with masses of $10^6 M_\odot, 10^7 M_\odot,$ and $10^8 M_\odot$ with scale radii of 162 pc, 512 pc, and 1.62 kpc respectively. As the Figure demonstrates, each subhalo creates gaps with a characteristic scale and this scale increases with the mass of the perturber. Thus, rather intuitively, the gap size is related to the mass of the subhalo which created the gap as was first pointed out in \cite{yoon_etal_2011}. We also see that even for low mass subhaloes with $M=10^6 M_\odot$, the characteristic gap size is $\sim 3^\circ$, while for those with $M=10^7 M_\odot$, the typical width is of order of $\sim10^\circ$.

\begin{figure}
\centering
\includegraphics[width=0.5\textwidth]{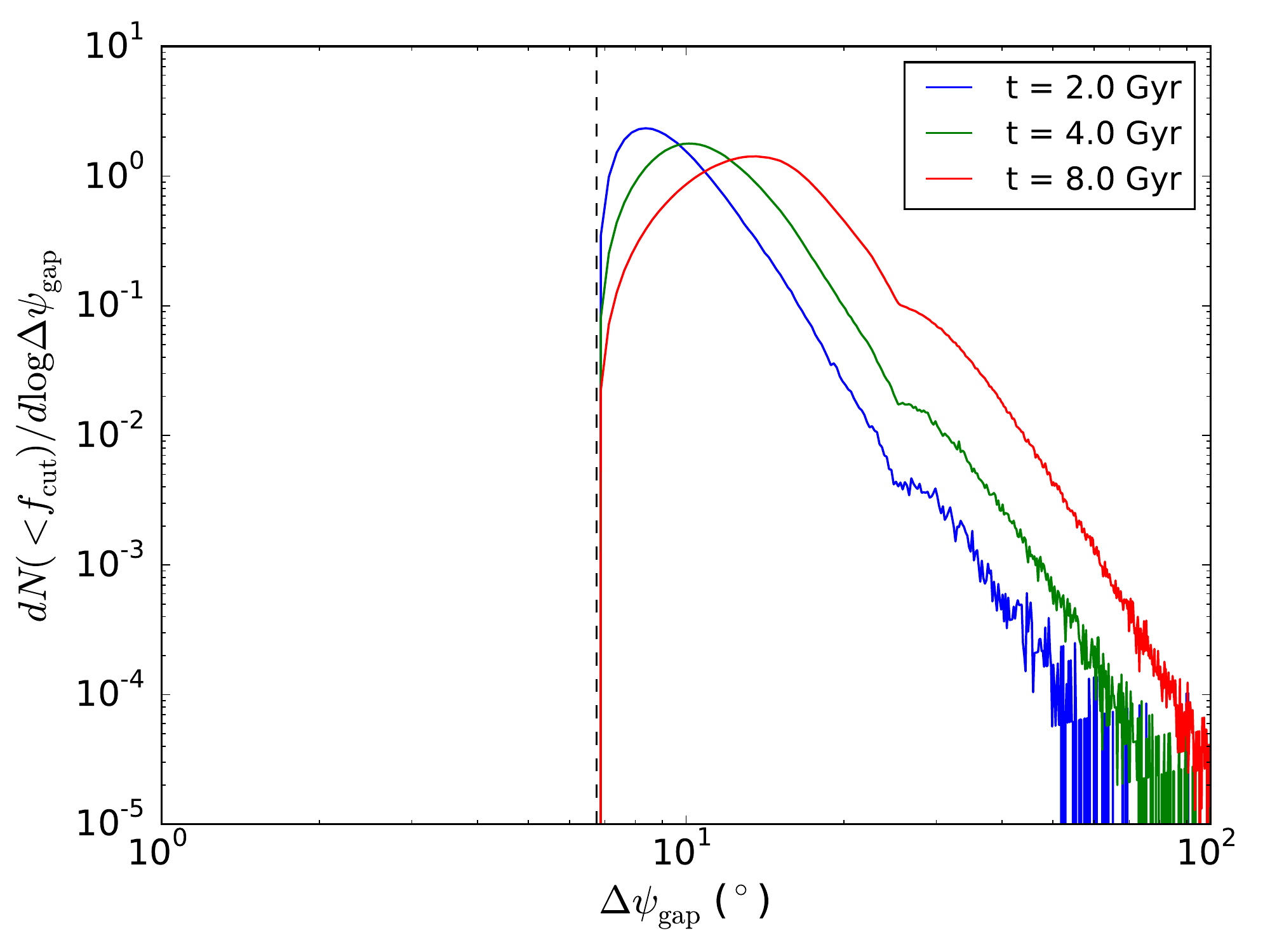}
\caption{Distribution of gap sizes created by a $M=10^7 M_\odot$ and $r_s = 512$ pc subhalo in streams of various ages which create gaps deeper than 50\%. The vertical dashed black line comes from \protect\eqref{eq:gap_size_thresh_grow}. Unsurprisingly, we see that for older streams, the gaps are larger since they have had more time to grow.} 
\label{fig:gap_size_dist_vary_time}
\end{figure}

Next, we consider the effect of changing the age of the stream. In \Figref{fig:gap_size_dist_vary_time}, we show the distribution of gap sizes created by subhaloes of the same mass, but with three different stream ages. We see that as the stream grows in age, so too do the sizes of the gaps in the stream. This is simply because the gaps in these streams had more time to expand. 

\begin{figure}
\centering
\includegraphics[width=0.5\textwidth]{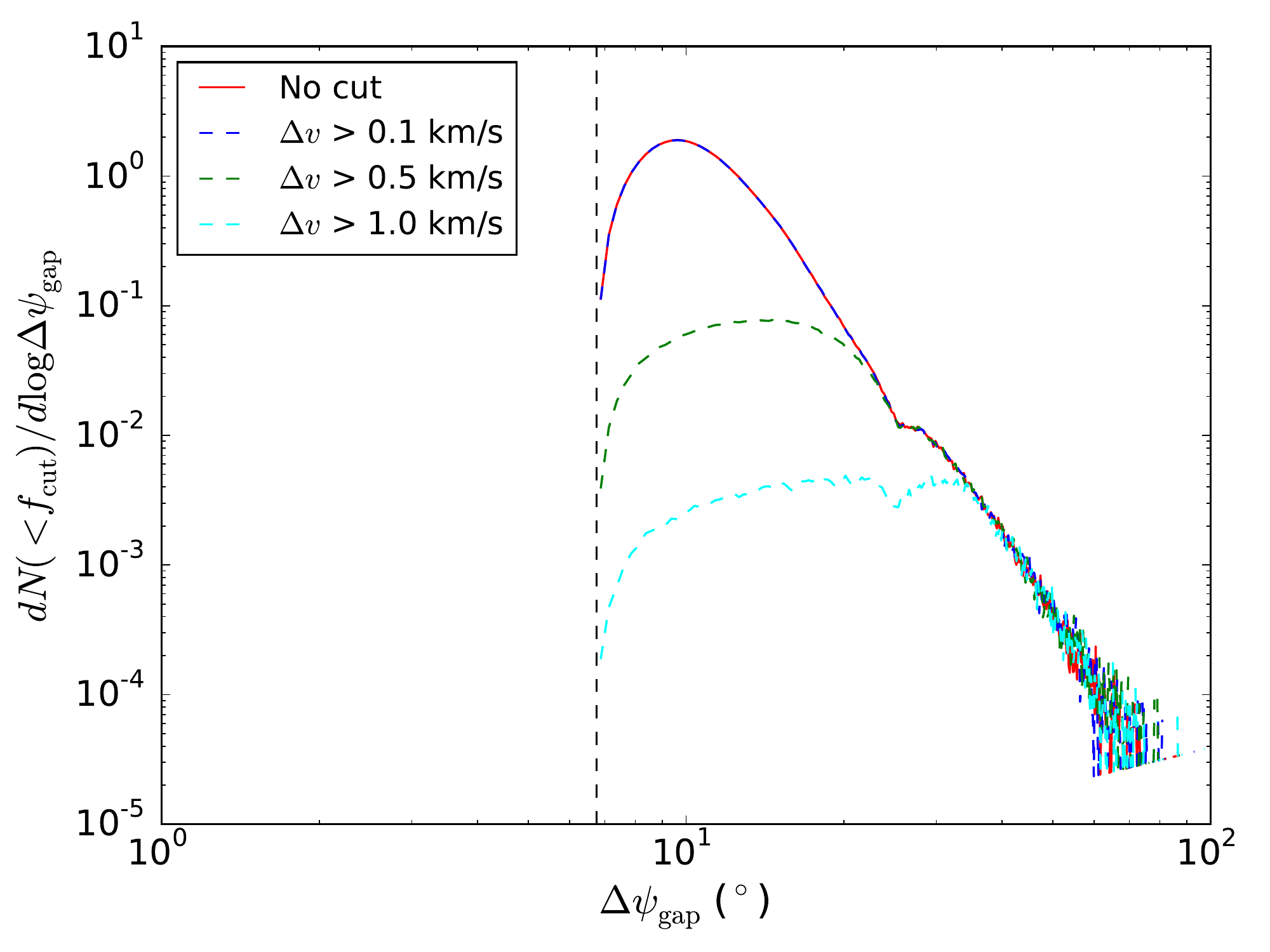}
\caption{Distribution of gap sizes for a subhalo with $M=10^7 M_\odot$ and $r_s = 512$ pc for gaps deeper than 50\%, i.e. $f<0.5$.  The colored lines correspond to different thresholds for the maximum velocity kick during the flyby. The vertical dashed black line comes from \protect\eqref{eq:gap_size_thresh_grow}. We note that the curve with no cut and the curve with $\Delta v > 0.1$ km/s are indistinguishable.} 
\label{fig:gap_size_dist_vcut}
\end{figure}

Finally, we explore how the gap size depends on the velocity cutoff we use. This cutoff specifies what the maximum velocity of a subhalo kick must be in order to be included. As we discussed in \Secref{sec:effective_nbody_vel_kicks}, our effective N-body simulations suggest that detectable density depletions exist for kicks with $\Delta v > 0.1$ km/s. In \Figref{fig:gap_size_dist_vcut} we show the gap size distributions for various $\Delta v$ thresholds. We see that requiring $\Delta v > 0.1$ km/s captures almost all of the visible kicks so our model is not missing very much in this example. We note that the gaps used in this figure were required to have $f<0.5$ and that for shallower gaps, a larger fraction of the kicks would be below the velocity kick threshold. 

As we will discuss in \Secref{sec:limitations}, our approach is based on circular orbits and does not account for the eccentricity of the stream. This eccentricity causes the gap size to oscillate with the galactocentric distance as $r^{-2}$, e.g. Fig. 13 of \cite{sanders_bovy_erkal_2015}. As a result, the distribution of gap sizes will depend on the exact phase at which the stream is measured and our results should be thought of as giving the average behavior of the gap size. These effects are beyond the scope of this work but are included in the recent work of \cite{bovy_erkal_sanders} who find broadly similar results by studying the power spectrum of the perturbed stream density and find the the majority of the power is on scales larger than 10$^\circ$.

\subsection{Gap depth for single subhalo population} \label{sec:gap_depth_dist_homogenous}

\begin{figure}
\centering
\includegraphics[width=0.5\textwidth]{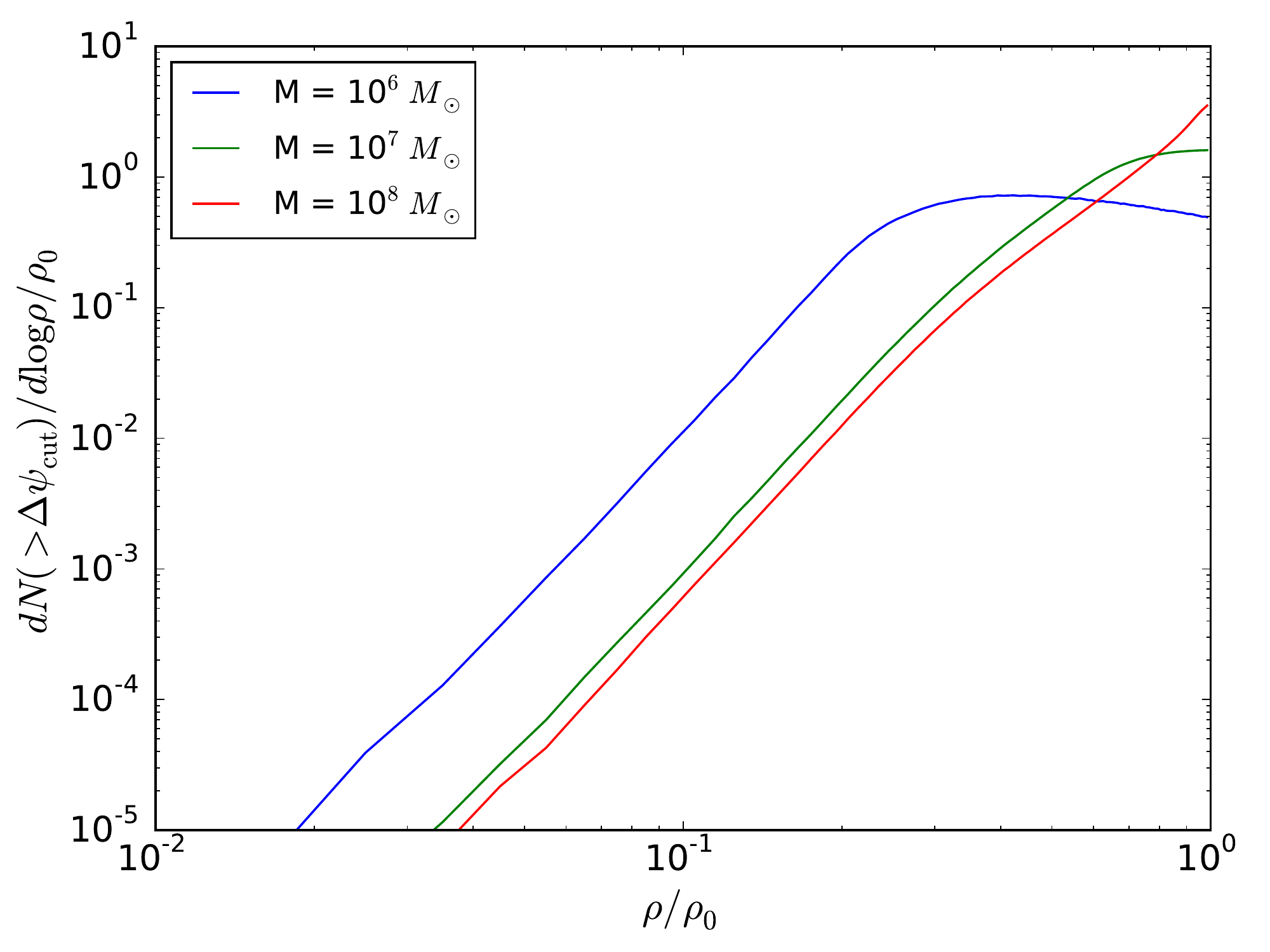}
\caption{Distribution of gap depths for a subhalo with varying mass and scale radius which create gaps larger than $1^\circ$. As the mass is decreased, the number of shallow gaps decreases while the behavior for deep gaps remains unchanged. This decrease is due to the requirement of a minimum velocity kick of $\Delta v > 0.1$ km/s.  } 
\label{fig:gap_depth_dist_mass}
\end{figure}

The detectability of the stream gap depends not only on its size, but
also on the density contrast between the center of the gap and the
unperturbed stream. This Section therefore looks at the gap depth
distribution. We repeat the same procedure as in
\Secref{sec:gap_size_dist_homogenous} and sample the appropriate
distributions for velocity components, impact parameter, and impact
time. \Figref{fig:gap_depth_dist_mass} shows the distribution of gap
depths imparted by three different mass subhaloes. As the mass of the
perturber is decreased, the number of shallow gaps decreases but the
number of deep gaps is unchanged. The decrease at the shallow end is
due to the imposition of a minimal velocity kick, which affects
the lower mass subhaloes more. The convergence for deep gaps is due
to the fact that for direct impacts, and hence the deepest gaps, the gap depth itself (i.e. \eqref{eq:gap_density})
does not depend on mass for the scaling between mass and scale radius
used here.

\begin{figure}
\centering
\includegraphics[width=0.5\textwidth]{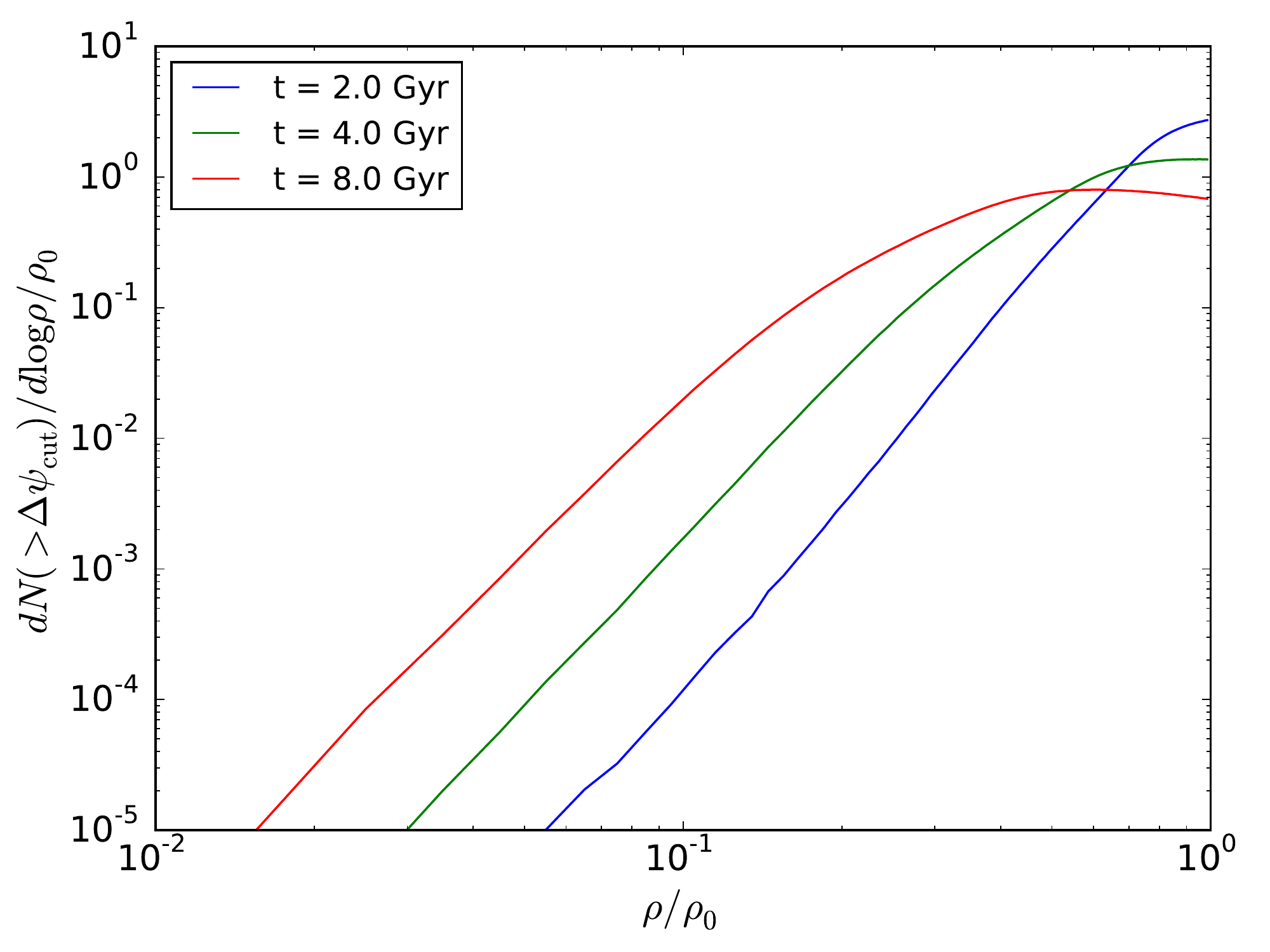}
\caption{Distribution of gap depths for a subhalo with a mass of $M=10^7 M_\odot$ and $r_s = 512$ pc in streams of various ages which create gaps larger than $1^\circ$.  As expected the younger streams have shallower gaps and the older streams have deeper gaps. } 
\label{fig:gap_depth_dist_time}
\end{figure}

In \Figref{fig:gap_depth_dist_time} we show the effect of changing the
age of the stream on the distribution of gap depths. As expected,
younger streams in which gaps have less time to grow have shallower
gaps and older streams have deeper gaps.

\subsection{Gap size distribution from $\Lambda$CDM background} \label{sec:gap_size_dist_lcdm}

\begin{figure}
\centering
\includegraphics[width=0.5\textwidth]{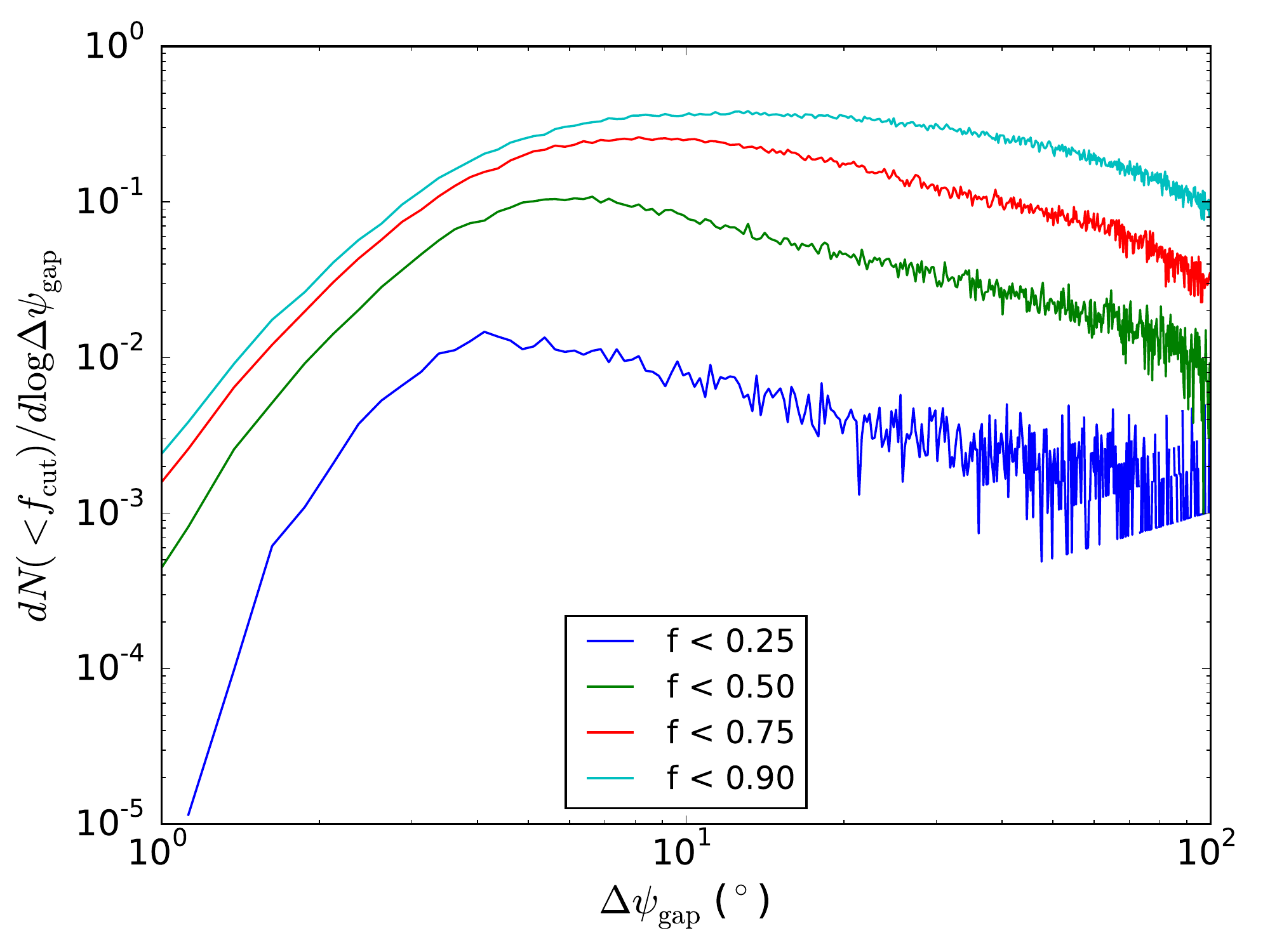}
\caption{Distribution of gap sizes for subhaloes with $10^5 M_\odot < M < 10^9 M_\odot$ for the Pal 5 stream. This distribution is not sensitive to masses below $10^5 M_\odot$ since most of the velocity kicks from those subhaloes are too small to create a visible gap. } 
\label{fig:gap_size_dist_LCDM}
\end{figure}

Having looked at the individual facets of the stream gap behaviour, we
combine the intuition gained in the previous sub-sections to analyse
the gap properties expected from a population of $\Lambda$CDM
subhaloes. We consider subhaloes with masses in the range $10^5-10^9
M_\odot$. Subhaloes below this mass range create too small of a
velocity kick to cause a significant density depletion. We repeat the
same procedure as in \Secref{sec:gap_size_dist_homogenous},
marginalizing over the flyby velocities, impact parameter, and impact
time, as well as marginalizing over the subhalo mass assuming
cosmologically motivated halo mass function described
above. \Figref{fig:gap_size_dist_LCDM} gives the expected distribution
of gap sizes these subhaloes would create for various gap density
thresholds. We see that the peak of the distribution depends on what
density threshold is used but for a feasible depth of $f<0.75$, the
characteristic scale is on the order of several degrees.

\begin{figure}
\centering
\includegraphics[width=0.5\textwidth]{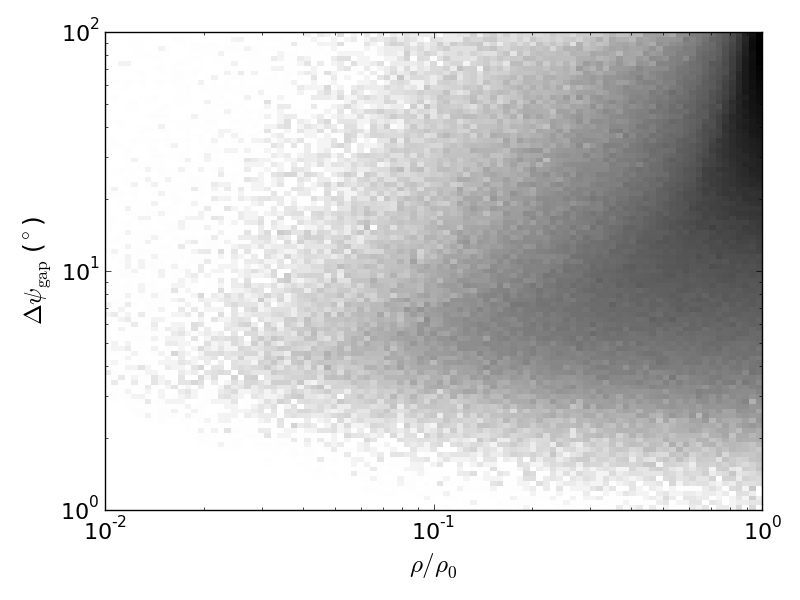}
\caption{Two dimensional distribution of gap depths and sizes for subhaloes with $10^5 M_\odot < M < 10^9 M_\odot$ for the Pal 5 stream with velocity kicks above $\Delta v > 0.1$ km/s.  For very shallow gaps, there is a larger range of gap sizes as seen in \protect\figref{fig:gap_size_dist_LCDM} which shows various slices of this figure. As we move towards deeper gaps, the typical gap size becomes smaller. This is because of the increasing influence of the lower mass haloes which create smaller gaps. } 
\label{fig:gap_depth_size_dist_LCDM}
\end{figure}

In \Figref{fig:gap_depth_size_dist_LCDM} we show the two dimensional distribution of gap sizes and depths created from a $\Lambda$CDM spectrum of subhaloes with masses between $10^5-10^9 M_\odot$. For shallow gaps, there are a wide range of gap sizes with the larger gap sizes dominating. However, as we proceed to deeper gaps, the lower mass subhaloes become more important which leads to smaller gaps. This somewhat counterintuitive result is due to the requirement of a minimum velocity kick. For deep gaps, this is satisfied for both low and high mass subhaloes and the gap size is dominated by the low mass subhaloes since they are more numerous. However, for shallow gaps, many of the low mass flybys will produce a negligible kick (e.g. \figref{fig:gap_depth_dist_mass}) so the shallow gaps have a larger contribution from larger subhaloes and hence larger gaps. 

\section{Number of gaps expected in observed MW streams} \label{sec:gaps_in_real_streams}

Now that we have explored the properties of gaps from a realistic population of subhaloes, we can compute these quantities for the stellar streams observed in the Milky Way. Here, we will focus on globular cluster streams since these have the smallest velocity dispersion and hence should be sensitive to the widest range of subhalo masses \citep[see][for a detailed discussion on the gap information content]{subhalo_properties}. There are at least 13 claimed globular cluster streams to date \citep{stream_book_grillmair_carlin}. In \Tabref{tab:obs_streams} we give the properties of 6 of these streams. These stream characteristics can be used to make a prediction for the expected number of gaps in each case. To proceed, the estimate of the progenitor mass is obtained from the model of the stream width described in \cite{stream_width}. Finally, the stream's age can be gleaned using a simple model for the stream growth rate which we will present below. Note that the fiducial stream used in \Secref{sec:gap_properties} was based on Pal 5 so \Figref{fig:gap_size_dist_LCDM} shows the distribution of gap sizes expected in Pal 5. 

\begin{table*}
\caption{Observed properties for six cold streams around the Milky Way. Pal 5 data: $^a$ \protect\cite{gd_pal5}, $^b$\protect\cite{pal5disc}, $^c$ \protect\cite{kuepper_et_al_pal5}. GD-1 data: $^d$ \protect\cite{gd1disc}, $^e$\protect\cite{bowden_et_al_gd1}. Acheron and Sytx data: $^f$\protect\cite{g_acheron_styx_cocytos_lethe}. Tri/Psc data: $^g$\protect\cite{bonacadisc}, $^h$\protect\cite{tripsc_pandas}, $^i$\protect\cite{tripsc_charlesmartin}. ATLAS data: $^j$\protect\cite{atlasdisc}. Phoenix data: $^k$\protect\cite{phoenix_disc}. Note that the ages for Styx, Tri/Psc, ATLAS, and Phoenix come from the stream length results in \protect\secref{sec:stream_age}. For the same four streams, the progenitor masses are estimated using their respective streams widths and the results of \protect\cite{stream_width}. We note that the inferred age of the Styx stream is greater than the age of the universe so we use an age of 13 Gyr when computing its gap properties. Note that we only have information on the pericenters and apocenters for Pal 5 and GD-1. For the other four streams these fields are left blank. }
\begin{tabular}{|c|c|c|c|c|c|c|c|c|c|}
\hline
Stream & $\Delta \phi_{\rm helio}$ & $\Delta \phi_{\rm GC}$ & $l_{\rm stream}$ & $r_{\rm helio}$ &$r_{\rm GC}$ & $r_{\rm apo}$ & $r_{\rm peri}$ & Age & Progenitor \\ \hline
Pal 5 & 22$^\circ$ $^a$& 29$^\circ$ & 9 kpc & 23 kpc $^b$ & 19 kpc $^b$& 19.5 kpc $^b$ & 5.7-7.4 kpc $^{b,c}$& 3.4 Gyr $^c$ & $2\times 10^4 M_\odot$ $^c$ \\ 
GD-1 & 63$^\circ$ $^d$& 32$^\circ$ & 9 kpc & 7.3-9.1 kpc $^d$& 13.5-15 kpc $^d$& 20-25 kpc $^e$ & 14 kpc $^e$& 3-5 Gyr $^e$ & $10^5 M_\odot$ $^e$\\ 
Tri/Psc & 35$^\circ$ $^{g,h}$& 17$^\circ$& 12 kpc & 35 kpc $^i$ & 40 kpc& & & 9.3 Gyr & $2 \times 10^4 M_\odot$\\ 
ATLAS & 20$^\circ$ $^{j}$& 16$^\circ$ & 6 kpc & 20 kpc $^j$ & 22-23 kpc& & & 2.1 Gyr & $1\times 10^5 M_\odot$\\ 
Phoenix & 15$^\circ$ $^{k}$ & 11$^\circ$ & 4 kpc & 17.5 kpc $^k$& 18-19 kpc& & & 1.8 Gyr & $4 \times 10^4 M_\odot$\\ 
Styx & 53$^\circ$ $^f$& 64$^\circ$ & 50 kpc & 45 kpc $^f$& 45-50 kpc& & & 18.5 Gyr & $3\times 10^5 M_\odot$ \\ 
\hline
\end{tabular}
\label{tab:obs_streams}
\end{table*}

\subsection{Estimating the age of a stream} \label{sec:stream_age}

As we saw in \Secref{sec:gap_size_dist_homogenous} and \Secref{sec:gap_depth_dist_homogenous}, the properties of gaps depend on the age of the stream. This is because gaps will have had more time to grow in older streams. In addition to changing the properties of the gaps, the age of the stream also determines how many interactions there will have been, as prescribed in \eqref{eq:num_encounters}. As such, in order to estimate the number of gaps in a stream we need to know the age of the stream. For some of the streams we consider here, dynamical modelling has already been performed using Lagrange point stripping methods, e.g. \cite{kuepper_et_al_pal5} for Pal 5 and \cite{bowden_et_al_gd1} for GD-1. This modelling gives an estimate of the age for both of these streams. However, for the other four streams there are no such estimates yet. 

\subsubsection{Estimating the progenitor mass} \label{sec:mass_estimate}

Let us use the observed width of the stream to estimate the mass of the progenitor. The evolution of the width of the stream perpendicular to the orbital plane was studied in \cite{stream_width} for both axisymmetric and triaxial potentials. Here, we will just use the result for spherical potentials that the stream width is given by
\eq{ w = \frac{1}{\sqrt{2}} \frac{\sqrt{\frac{Gm}{3 r_{\rm tidal}} }}{v_{\rm peri}} , }
where $m$ is the mass of the progenitor, $r_{\rm tidal}$ is the tidal radius, and $v_{\rm peri}$ is the velocity of the progenitor at pericenter. For the streams whose orbits we do not know, we will assume they are on circular orbits and the orbital velocity is given by
\eq{ v = \sqrt{\frac{GM(<r)}{r}} ,} 
where $M(<r)$ is the mass of the host potential enclosed within a radius of $r$. For a progenitor on a circular orbit, the tidal radius is given by
\eq{ r_{\rm tidal} = r \Big( \frac{m}{(4-\gamma^2) M(<r)} \Big)^{\frac{1}{3}}  , }
where $\gamma$ is given by \eqref{eq:gamma}. 
For a host galaxy with a flat rotation curve, i.e. a logarithmic potential, we get $\gamma^2 = 2$. Plugging these into the formula for the width we see
\eq{ w = \frac{1}{\sqrt{6}} \Big( \frac{m}{2 M(<r)} \Big)^{\frac{1}{3}} . \label{eq:width}}
Finally, this can be arranged to give
\eq{ m = 2^{\frac{1}{2}} 3^{\frac{3}{2}} w^3 M(<r) . }
Using the widths of these streams reported in \cite{stream_width} we give the estimates of their progenitor masses in \Tabref{tab:obs_streams}.

For Pal 5, we assume a circular velocity of $220$ km/s at a radius of $13$ kpc, as measured in \cite{bovy_et_al_2012}, and get a progenitor mass of $2.3 \times 10^4 M_\odot$, very similar to that reported in \cite{kuepper_et_al_pal5}. For GD-1, we assume a circular velocity of $220$ km/s at a radius of $19$ kpc and get a mass of $1.2 \times 10^4 M_\odot$, which is below the range of masses considered in \cite{bowden_et_al_gd1} although they have broad broad posteriors. Instead of this estimate, we use the central value from \cite{bowden_et_al_gd1} of $10^5 M_\odot$ but we will also discuss the gap predictions if GD-1 is as low mass as our method suggests. For Styx, we assume a circular velocity of $190$ km/s at a radius of $45$ kpc, as measured by \cite{deason_et_al_2012}, and get a mass of $2.6 \times 10^5 M_\odot$. For Tri/Psc, we assume a circular velocity of $190$ km/s at a radius of $40$ kpc and get a mass of $2.4\times 10^4 M_\odot$. For ATLAS, we assume a circular velocity of $220$ km/s at a radius of $22$ kpc and get a mass of $1.3 \times 10^4 M_\odot$. Finally, for Phoenix we assume a circular velocity of $220$ km/s a radius of $19$ kpc and get a mass of $3.8 \times 10^4 M_\odot$. 

\subsubsection{Estimating the stream age}

With the progenitor mass in hand, we can now estimate the age of a stream. Once again, the stream is assumed to follow a circular orbit. The length of a stream can be estimated by considering small perturbations to the circular orbit of the progenitor and determining how quickly these stars move away from the release point. This is similar to the analysis presented in \cite{three_phases}, where the effect of changing the velocity is considered. However, we must also include the fact that stripped material will be offset from the progenitor in radius, i.e. it is released from the Lagrange points. Thus, we find that the stream length in radians grows as
\eq{ l \sim 2 \frac{(4-\gamma^2)^{\frac{2}{3}}}{\gamma^2}  \Big( \frac{m}{M} \Big)^{\frac{1}{3}} \frac{v t}{r} , \label{eq:stream_length}} 
where $v$ is the circular velocity at the radius $r$. We note that similar expressions have been derived in \cite{amorisco_2014} (see their equation 20). \eqref{eq:stream_length} can then be re-arranged to give an estimate for the age of the stream in terms of its length and mass, as well as the mass of the host enclosed within that radius. As above, we assume that the potential has a flat rotation curve so $\gamma^2 = 2$, and turn \eqref{eq:stream_length} into
\eq{ t = \frac{2^{-\frac{2}{3}} l }{ \Big(\frac{m}{M}\Big)^{\frac{1}{3}} \sqrt{\frac{GM}{r^3}} } . \label{eq:age} }
For each of the streams, we give an estimate of their age in \Tabref{tab:obs_streams}. For Pal 5 we get an age of 3.4 Gyr which matches \cite{kuepper_et_al_pal5}. For GD-1, if we use a mass of $10^5 M_\odot$ we get an age of 3.8 Gyr, in the center of the range found in \cite{bowden_et_al_gd1}. We will use a value of 4 Gyr but we note that if we instead use our inferred mass, we would get an age of 7.7 Gyr, making GD-1 substantially older than previous fits have suggested. We note that the estimates of the progenitor mass and stream age in this section relied on results from circular orbits, however stream modelling for more realistic streams on eccentric orbits can also be used to estimate these quantities more robustly \citep[e.g.][]{gibbons_et_al_2014,bovy_2014,sanders2014,kuepper_et_al_pal5,fardal2015,bowden_et_al_gd1}. With these ages, we can now estimate the number and properties of the gaps in these streams.

\subsection{Pal 5}

We can now make tailor-made predictions for each of the six streams reported in Table~\ref{tab:obs_streams}, starting with Pal 5. Based on \Tabref{tab:obs_streams}, we will model Pal 5 as being on a circular orbit with a radius of 13 kpc. The number density of subhaloes between $10^6-10^7 M_\odot$ at this radius is $1.01 \times 10^{-3} {\rm kpc}^{-3}$. Assuming a flat rotation curve with a circular velocity of 220 km/s found in \citep{bovy_et_al_2012}, we get an orbital period of $360$ Myr. \Figref{fig:gap_size_dist_LCDM} shows the distribution of gap sizes in Pal 5 from subhaloes with a mass in the range $10^5 M_\odot - 10^9 M_\odot$. It predicts that Pal 5 should have a characteristic gap size of $\Delta \psi_{\rm gap} \sim 4-7^\circ$ for deep gaps with $f<0.5$ but the distribution is quite broad. This also guides the scales on which one should be looking for gaps in Pal 5.

\begin{figure}
\centering
\includegraphics[width=0.5\textwidth]{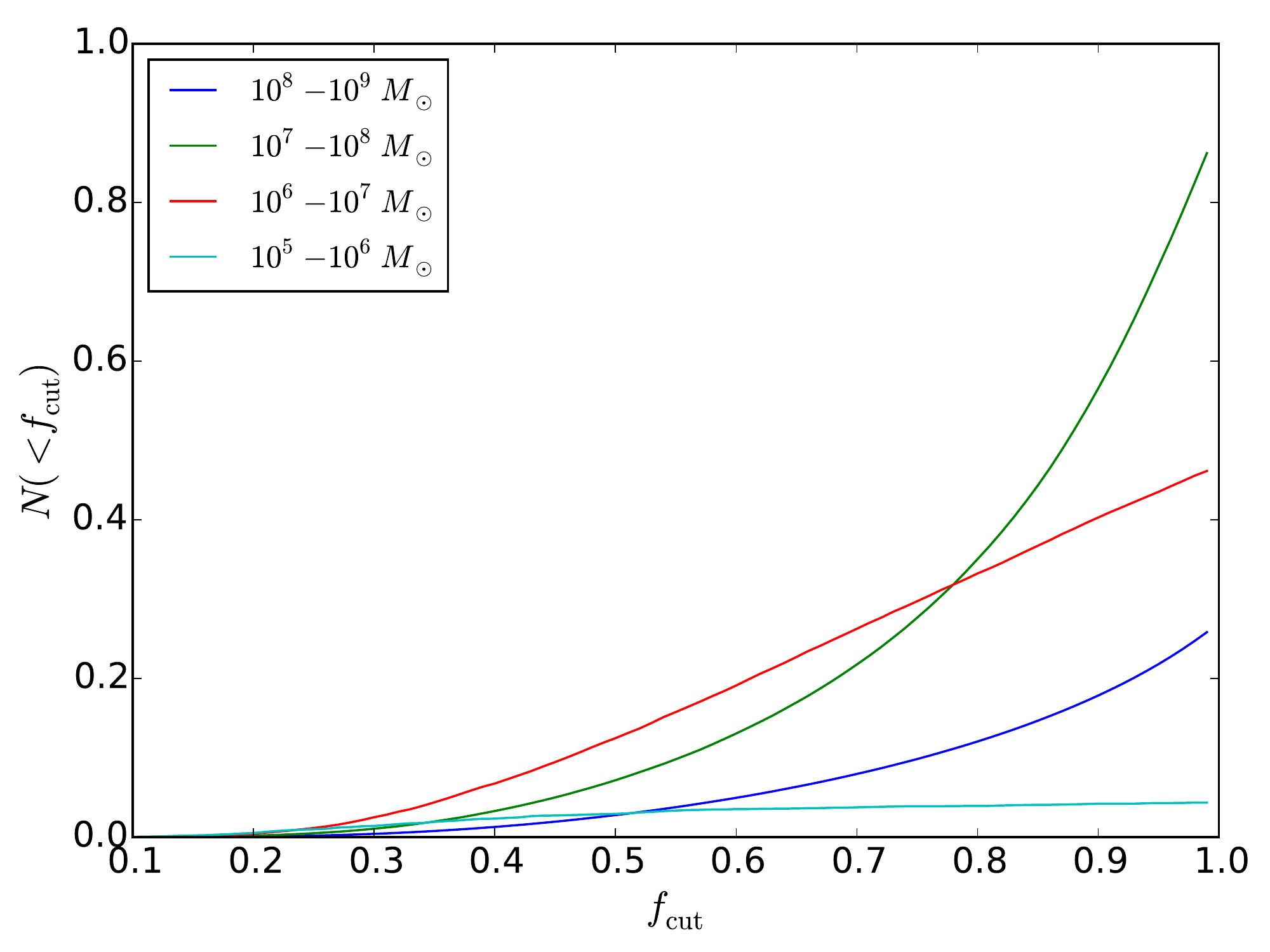}
\caption{Expected number of gaps deeper than $f_{\rm cut}$ with gap sizes bigger than 1$^\circ$ in Pal 5. The different colored curves show the expected number of gaps in different mass ranges. We see that gaps from subhaloes with masses between $10^6-10^7 M_\odot$ dominate for most density thresholds. Combining all the mass ranges together, Pal 5 should have $\sim$0.7 gaps with densities less than 0.75.  } 
\label{fig:expected_gaps_Pal5}
\end{figure}

In addition to the distribution of gap sizes, we can also compute the expected number of gaps. We do this by performing the same marginalization over the flyby velocities, impact parameter, impact time, and subhalo mass described in \Secref{sec:gap_size_dist_lcdm}. We include all flybys within 5 kpc of the stream and only find a 1-2\% change in the number of gaps if we extend this to 10 kpc, suggesting that our results have converged. In \Figref{fig:expected_gaps_Pal5} we show the expected number of gaps in Pal 5 deeper than a given threshold. We see that over a range of gap density thresholds, the subhaloes from $10^6 M_\odot-10^8 M_\odot$ produce the dominant contribution. Above a gap density threshold of $f_{\rm cut} \sim 0.75$, gaps from the subhaloes with a mass of $10^7 M_\odot - 10^8 M_\odot$ outnumber those from the $10^6-10^7 M_\odot$ range. We see that if we take $f_{\rm cut} = 0.5$ we expect to find 0.3 gaps in Pal 5. However, if we take $f_{\rm cut} = 0.75$, we would expect almost 0.7 gaps. Thus, the null detection reported in \cite{ibata_et_al_pal5} is not be very surprising although as noted above, the search was performed on scales significantly smaller than described here. The prediction for the number of gaps in Pal 5, and the other five streams, is given in \Tabref{tab:num_gaps}. 

\begin{table}
\begin{center}
\caption{Expected number of gaps and their characteristic size for six observed cold streams ranked by the number of gaps. The second and fourth column respectively give the number of gaps deeper than 50\% and 75\% of the unperturbed density. The third and fifth column give the most common gap size for gaps deeper than 50\% and 75\% of the unperturbed density. Styx has by far the most expected gaps and if it is a cold stream with the reported length, it will be the best candidate for finding subhaloes. Tri/Psc, Pal 5, and GD-1 all have a relatively similar number of gaps we can expect at least one deep gap with $f<0.5$ amongst them. These predictions include the gaps created from subhaloes with masses in the range $10^5 M_\odot < M < 10^9 M_\odot$ and account for the factor of 3 depletion expected from the presence of the MW disk \protect\citep{donghia_et_al_2010}. }
\begin{tabular}{|c|c|c|c|c|}
\hline
Stream & $N(<0.5)$ & $\Delta \psi_{\rm char}(<0.5)$ & $N(<0.75)$& $\Delta \psi_{\rm char}(<0.75)$  \\ \hline
Tri/Psc & 0.9 & 4$^\circ$ & 1.6 & 4.5$^\circ$\\
Pal 5 & 0.3 & 6$^\circ$ & 0.7 & 8$^\circ$\\ 
GD-1 & 0.3 & 5$^\circ$ & 0.6 & 6.5$^\circ$\\
ATLAS & 0.02 & 3$^\circ$ & 0.1 & 4$^\circ$\\
Phoenix & 0.01 & 2.5$^\circ$ & 0.06 & 4$^\circ$\\
Styx & 6 & 6.5$^\circ$ & 9 & 10$^\circ$\\
\hline
\end{tabular}
\label{tab:num_gaps}
\end{center}
\end{table}

\subsection{GD-1}

GD-1 is modeled as being on a circular orbit with a radius of 19 kpc. The number density of subhaloes between $10^6-10^7 M_\odot$ at this radius is $8.66 \times 10^{-4} {\rm kpc}^{-3}$. Assuming a circular velocity of 220 km/s, we get an orbital period of 530 Myr. Performing the same sampling as for Pal 5, we find 0.3 and 0.6 gaps in GD-1 for $f<0.5$ and $f<0.75$ respectively. This is roughly the same as the prediction for Pal 5. Since the age and length of GD-1 and Pal 5 are similar, the distribution of gap depths will look very similar to \Figref{fig:expected_gaps_Pal5}. However, if we instead use the age of 7.7 Gyr we found in \Secref{sec:mass_estimate}, we would expect 0.9 and 1.7 gaps for $f<0.5$ and $f<0.75$ respectively. 

\subsection{Tri/Psc}

Next we model the Tri/Psc stream. We take it to be on a circular orbit at 40 kpc. The number density of subhaloes between $10^6-10^7 M_\odot$ at this radius is $5.51 \times 10^{-4} {\rm kpc}^{-3}$. Taking a circular velocity of 190 km/s \citep{deason_et_al_2012}, we get an orbital period of 1.3 Gyr. The number of gaps in Tri/Psc is 0.9 and 1.6 for $f<0.5$ and $f<0.75$ respectively. Thus, Tri/Psc is a better candidate in the search for gaps than GD-1 or Pal 5. Furthermore, Tri/Psc is sufficiently far away from the disk that the factor of 3 depletion may be an overestimate. \cite{donghia_et_al_2010} found that the depletion is the strongest near the disk and decreases as we move away from the disk. 

\begin{figure}
\centering
\includegraphics[width=0.5\textwidth]{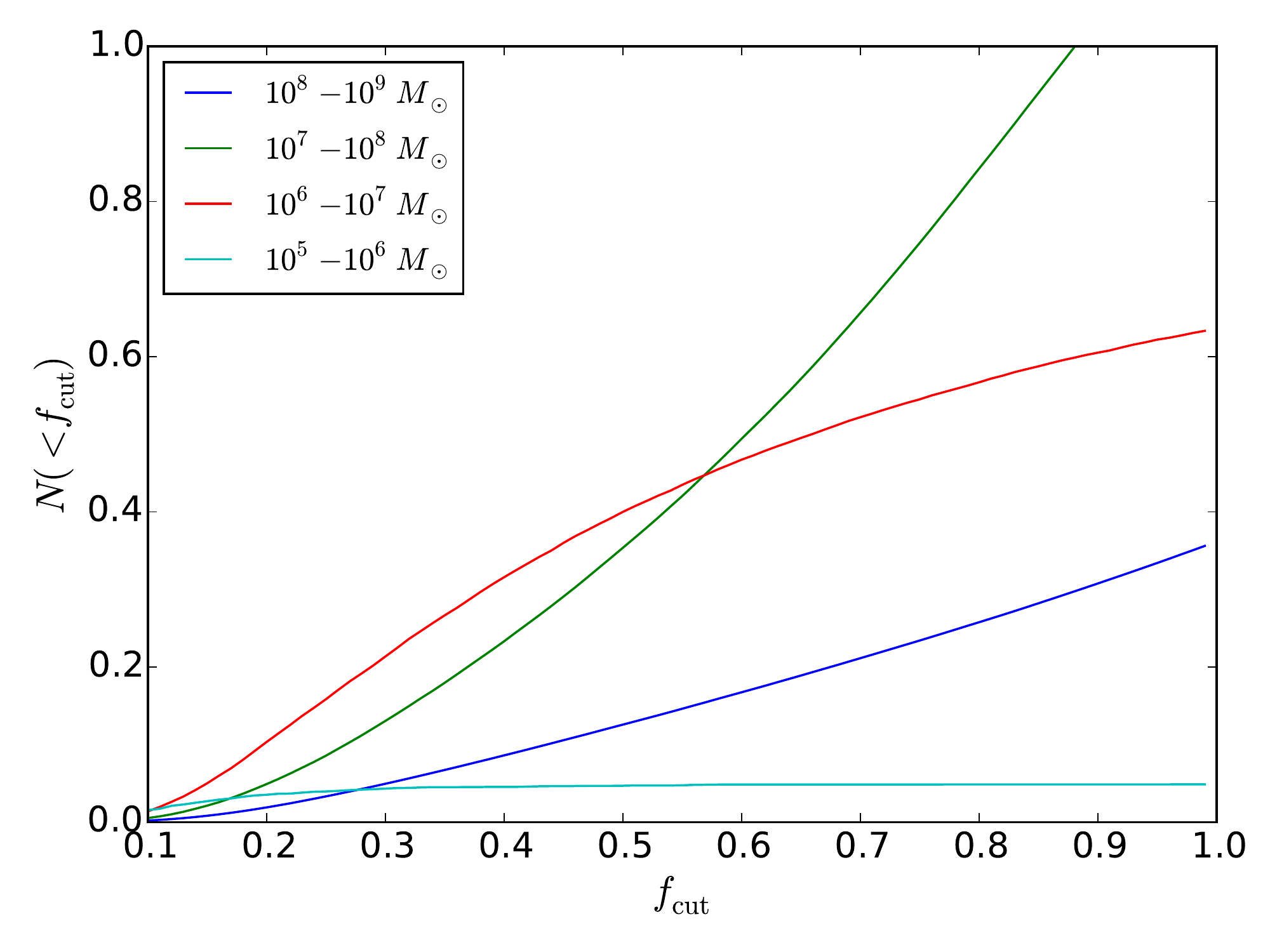}
\caption{Expected number of gaps deeper than $f_{\rm cut}$ with gap sizes bigger than 1$^\circ$ in Tri/Psc. The different colored curves show the number of gaps in various mass ranges. As with Pal 5, we see that subhaloes in the range $10^6-10^8 M_\odot$ produce the most subhaloes. } 
\label{fig:expected_gaps_tripsc}
\end{figure}

\Figref{fig:expected_gaps_tripsc} presents the cumulative distribution of gap depths for Tri/Psc. This can be compared against \Figref{fig:expected_gaps_Pal5} where we show the distribution for Pal 5. Since Tri/Psc is both longer and older than Pal 5, there are more gaps. In addition, the increased age of Tri/Psc gives the gaps more time to grow and hence it has a larger fraction of deep gaps. This can also be seen in \Figref{fig:gap_depth_dist_time} where the effect of the stream age on the distribution of gap depths was examined and the same result was found. 

\subsection{ATLAS}

We model the ATLAS stream as being on a circular orbit with a radius of 22 kpc. The number density of subhaloes between $10^6 - 10^7 M_\odot$ at this radius is $8.06 \times 10^{-4} {\rm kpc}^{-3}$. Taking a circular velocity of 220 km/s, we get an orbital period of 610 Myr. The expected number of gaps in the ATLAS stream is 0.02 and 0.1 for $f<0.5$ and $f<0.75$ respectively. This small number of gaps is due to the young age of ATLAS and the fact that the currently observed stream is quite short. Due to its young age, ATLAS will have shallower distribution of gaps than Pal 5.

\subsection{Phoenix}

Phoenix is modelled as being on a circular orbit with a radius of 19 kpc. The number density of subhaloes between $10^6-10^7 M_\odot$ at this radius is $8.66 \times 10^{-4} {\rm kpc}^{-3}$. Assuming a circular velocity of 220 km/s, we get an orbital period of 530 Myr. The number of gaps in Phoenix is 0.01 and 0.06 for $f<0.5$ and $f<0.75$ respectively. As with ATLAS, this is due to the young age of Phoenix and its short length. As a result, neither Phoenix nor ATLAS appear to be good candidates for detecting gaps. 

However, we note that \cite{grillmair_carlberg_2016} have recently suggested that the Phoenix stream may be part of a significantly longer stream which includes the Hermus stream \citep{grillmair_hermus_hyllus}. The purported length of 76 kpc would make it the longest cold stream in the Milky Way and hence an ideal candidate for studying gaps. Since the increased length of a stream also increases its estimated age, the expected number of gaps increases roughly quadratically with the stream's length. Thus, if the estimated length is correct, the combined stream would have approximately 20 gaps deeper than $f<0.75$.

\subsection{Styx}

The Styx stream \citep{g_acheron_styx_cocytos_lethe} is substantially longer than the other streams. If the stream is indeed a cold stream, then it is one of the best candidates for detecting subhaloes. We model Styx as being on a circular orbit with a radius of 45 kpc and take its age to be 13 Gyr, younger that the estimate in \Tabref{tab:obs_streams} but consistent with the age of the universe. The number density of subhaloes between $10^6-10^7 M_\odot$ at this radius is $5.01 \times 10^{-4} {\rm kpc}^{-3}$. Assuming a circular velocity of 190 km/s, we get an orbital period of 1.46 Gyr. Sampling over the flybys we find 6 and 9 gaps expected with $f<0.5$ and $f<0.75$ respectively. Thus, Styx could have an order of magnitude more gaps than the second best stream, Tri/Psc. However, \citep{g_acheron_styx_cocytos_lethe} argued that Styx is a dwarf galaxy stream. If this is correct, the number of gaps will decrease since the stream is younger and the stream from a dwarf galaxy is substantially hotter which will mask out many of the expected gaps.

\section{Numerical Testing} \label{sec:testing}

The framework we have discussed here has mostly been based on analytic methods, using the gap size formulae from \cite{three_phases} and the number of properties of the flybys as derived in \Secref{sec:flyby_properties}. In this section we will test these assumptions and see how well they work. 

\subsection{Flyby properties} \label{sec:test_flyby_properties}

First we compare the properties of the subhalo flybys. We take the N-body simulation of Pal 5 described in \Secref{sec:effective_nbody_vel_kicks} and include three times the expected subhalo population between $10^5-10^6 M_\odot$ using the fits from the number density profiles in Aquarius \citep[][]{springel_et_al_2008} as described in \Secref{sec:num_density_subhaloes}. We include three times the expected population to improve the statistics since the total number of flybys is not that large. Note that we also have not decreased the number density of subhaloes by a factor of 3 due to the effect of the MW disk so in reality, this example has roughly 9 times the expected subhaloes. These subhaloes are included as tracer particles which are sourcing the force expected if they were Hernquist profiles with their given mass and scale radius. The simulation is run for 3.4 Gyr which is the best fit age from \cite{kuepper_et_al_pal5}. Despite including significantly more substructure than expected, we see only a minor effect in the density from these low mass subhaloes. 

While the simulation is running, at each timestep we record whenever a subhalo passes within 2 kpc of a stream particle and record the position and velocity of both the subhalo and stream particle. During the simulation 2668 subhaloes passed within 2 kpc. We repeat our analysis above based on the stream length in the simulation at the present time and estimate that 2320 subhaloes should pass within 2 kpc of the stream in this time. Thus even though the model was based on circular orbits, it gives a good estimate for the number of flybys for a realistic stream on an eccentric orbit. 

\begin{figure}
\centering
\includegraphics[width=0.5\textwidth]{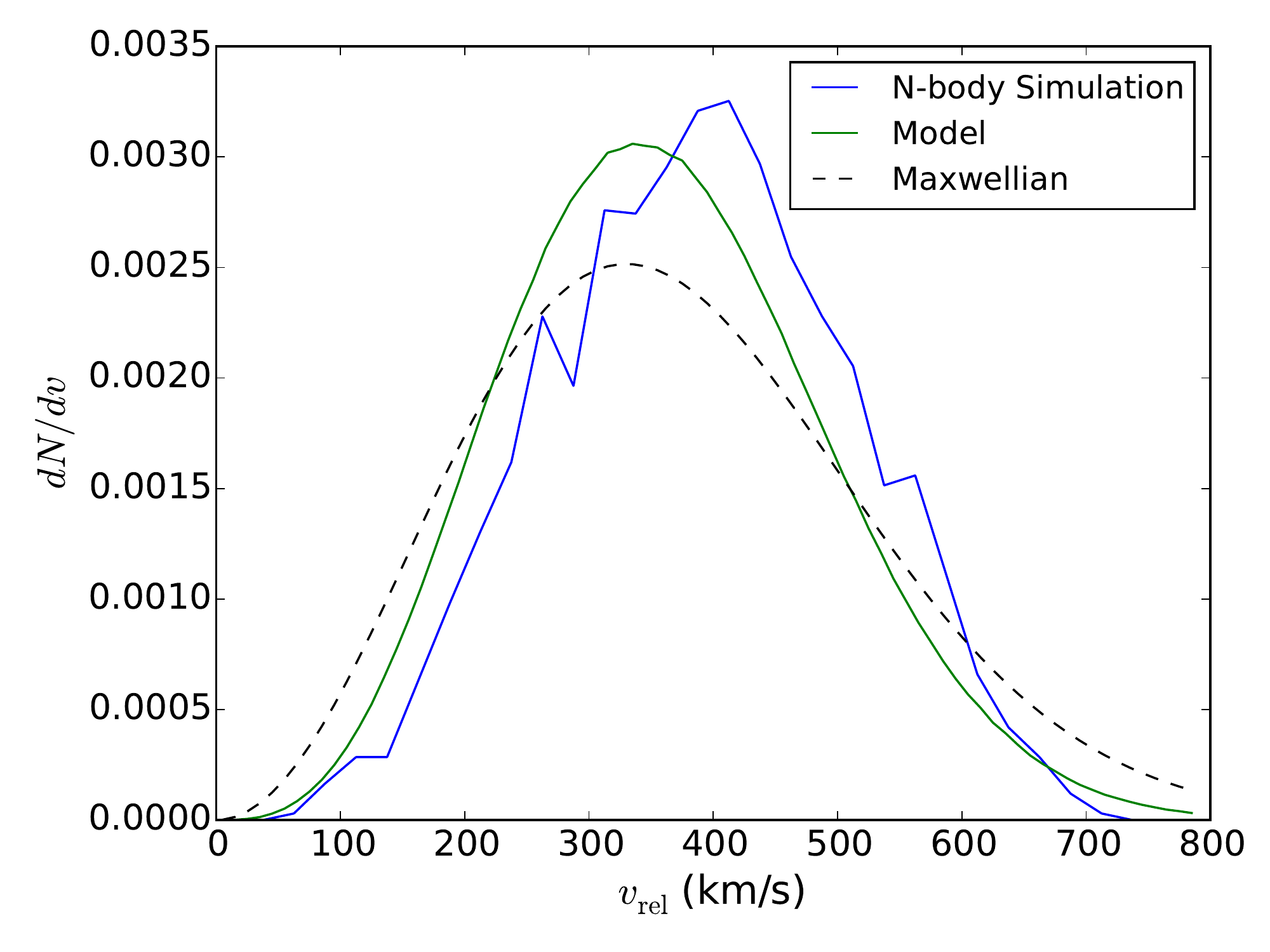}
\caption{Distribution of relative velocities between the stream and the subhalo flyby. In solid blue is the result of an N-body simulation of a Pal 5-like stream described in the text. In solid green is the model in this work from \protect\eqref{eq:Ppara} and \protect\eqref{eq:Pperp}. In dashed black is the relative velocity of particles in a Maxwellian distribution which was used in \protect\cite{yoon_etal_2011}. While the agreement is not as good as in \protect\figref{fig:vdist_flyby}, our model roughly matches the distribution in the simulation. } 
\label{fig:flyby_dist}
\end{figure}

In \Figref{fig:flyby_dist} we examine the distribution of relative flyby velocities for subhaloes which pass within 2 kpc. For each flyby in the simulation, we find the stream particle which has the smallest impact parameter and use the relative velocity to the subhalo. The figure also shows the prediction of our model and the relative velocity distribution from \cite{yoon_etal_2011}. We see that our model is a better match than using the relative velocity in a Maxwellian distribution. However, our model is not a perfect match because it does not account for the eccentricity of the stream. 

\subsection{Distribution of gap depths and sizes} \label{sec:eff_test}

\begin{figure}
\centering
\includegraphics[width=0.5\textwidth]{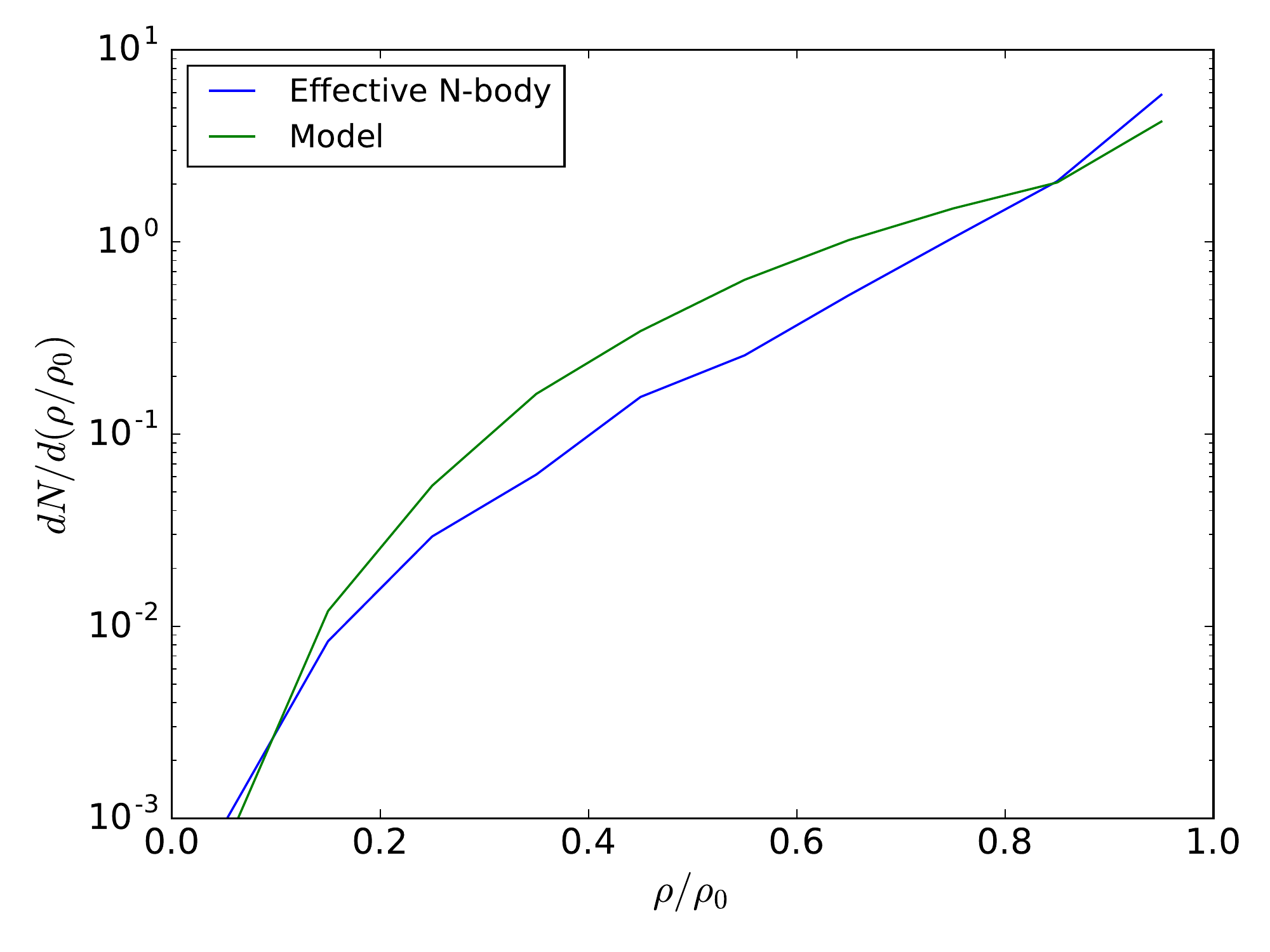}
\caption{Normalised distribution of gap depths in effective N-body simulations compared against the model. The blue curve shows the distribution of gap depths for a wide range of flybys simulated using the effective N-body method. The green curve shows the distribution of gap depths from our model. The match is quite good suggesting that the analytic model provides a reasonable estimate of the gap depth. The model predicts slightly deeper gaps than in the simulations which is likely due to our neglecting the dispersion of the stream in energy and angular momentum and the eccentricity of Pal 5's orbit.} 
\label{fig:dndrho}
\end{figure}

In addition to testing our assumptions about the flyby properties, we can also test how well our formalism works for determining the gap depths. We take the stream described in \Secref{sec:test_flyby_properties} and use the effective N-body formalism described in \Secref{sec:effective_nbody_vel_kicks} to simulate a large number of impacts. The properties of these impacts were chosen to cover a large range in the parameter space as follows: each component of the subhalo velocity was chosen uniformly from -500 to 400 km/s in steps of 100 km/s relative to the host potential, the impact parameter was chosen uniformly from 0 to 1 kpc in steps of 100 pc, five different impact times are chosen as 0.5, 0.85, 1, 1.13, 2 Gyr, and finally the subhalo mass is chosen uniformly in log space from $10^5 M_\odot$ to $1.024 \times 10^8 M_\odot$ in steps of $\log_{10}(2)$. The position of the impact along the stream was chosen to be halfway between the progenitor and the end of the stream. Each subhalo is modelled as a Plummer sphere with a scale radius given by \eqref{eq:plummer_r}. For each of these samples the effective N-body simulation is run, resulting in 500,000 simulated flybys. For each flyby, the particles are binned into $0.1^\circ$ bins and the density is saved for the region within 20$^\circ$ of the gap center. This density is then divided by the unperturbed density to get the gap depth and size. In addition, we use our model to make a prediction of the gap depth. For the model, we assume the stream is on a circular orbit at 15 kpc and compute the gap properties described in \Secref{sec:method}. In \Figref{fig:dndrho} we compare the distribution of gap depths between the effective N-body simulation with that of our analytic model and find a good match. There is a slight discrepancy in that the model predicts deeper gaps and in \Secref{sec:limitations} we will discuss the limitations of our model which are likely responsible for these differences. However, the match in \Figref{fig:dndrho} suggests that for a wide range of subhalo flybys, the analytic model used in this work produces a reasonable estimate of the actual gap depth. 

\begin{figure}
\centering
\includegraphics[width=0.5\textwidth]{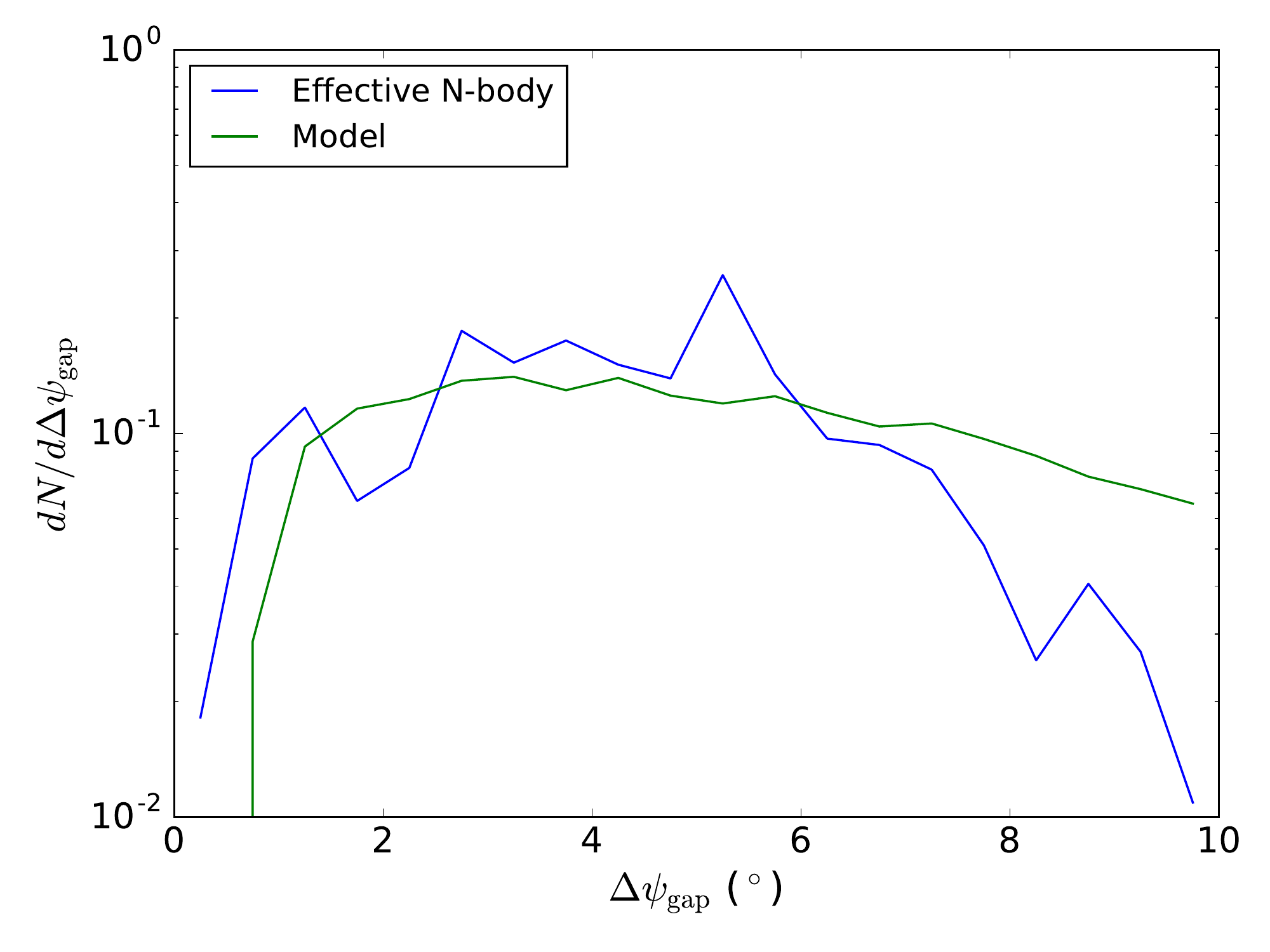}
\caption{Distribution of gap sizes in effective N-body simulations compared against the model. The blue curve shows the distribution of gap sizes for a wide range of flybys simulated using the effective N-body method. The green curve shows the distribution of gap sizes from our model. The match is quite good suggesting that the analytic model provides a reasonable estimate of the gap size as well. As noted in the text, we have restricted this comparison to gaps whose length is less than 10$^\circ$ in the effective N-body simulation since we cannot measure gaps which extend outside of the range of the unperturbed stream.  } 
\label{fig:dndpsi}
\end{figure}

Similarly, we can compute the distribution of gap sizes using the effective N-body simulations. Using the same distribution of flybys above, we show the distribution of gap sizes in \Figref{fig:dndpsi}. As with the gap depths, we get a similar distribution suggesting that our model is reproducing the gap size for a larger range of parameters. We also see that in the effective N-body simulation, the characteristic gap size is on the order of a few degrees. For this comparison we only considered flybys in the effective N-body simulation which created gaps less than 10$^\circ$ in size. This is because we only record the density profile within 20$^\circ$ of the gap center and the tails of the unperturbed stream only have lengths of roughly 20$^\circ$. Since we define the gap size by dividing the perturbed density by the unperturbed density and finding the size of the underdense region, we cannot find gaps which are longer than the length of the stream. 

\section{Discussion} \label{sec:discussion}

\subsection{Searching for gaps}

The results of this work can be used to guide targeted searches for stream gaps. In \Secref{sec:gap_properties}, the gap size distribution was explored for various mass subhaloes, and in \Figref{fig:gap_size_dist_LCDM} the distribution of gap sizes was shown for a $\Lambda$CDM population. The characteristic size of the gaps in known streams will depend on the age of the stream, but as we can see from \Figref{fig:gap_size_dist_LCDM}, it peaks between $4-8^\circ$ for a wide range of density thresholds, with a large number of gaps with sizes of order $10^\circ$ and above. In addition, the distribution drops off as we proceed to smaller gaps so gap searches should be performed on a scale of at least $1-2^\circ$. In recent work, \cite{ibata_et_al_pal5} searched for gaps in Pal 5 on the scale of $0.2-1^\circ$ and found none. The results of this work suggest that future searches will be more fruitful if they are made on larger scales. However, we stress that the gap sizes we found are as viewed from the center of the Galaxy. A heliocentric observer may see smaller or larger gaps depending on their distance to the stream. In addition, the gap may appear foreshortened if it is not oriented perpendicular to the line of sight. Thus while we have identified a characteristic size of gaps, these observational effects, as well as the stretching and compressing of gaps due to eccentric orbits discussed in \Secref{sec:limitations}, can modify the distribution of sizes depending on the exact orientation of the stream and its orbital phase.

In \Secref{sec:gaps_in_real_streams} we explored the number of gaps expected in known streams around the Milky Way. The majority of these gaps are quite shallow, with $\rho/\rho_0 > 0.5$. Thus, searches for gaps should be looking for shallow gaps over the scales of several degrees. This will require an accurate measurement of the stream density on fairly large scales. The detection of a wide and shallow gap will also require a careful modeling of the unperturbed stream density since changes in the stripping rate can also create features in the stream density profile.  

In \Tabref{tab:num_gaps} we give the expected number of gaps for six cold streams around the Milky Way. We find that if Styx is a genuine cold stream (i.e. originating from a globular cluster), it is by far the best candidate with 6 gaps expected with gaps deeper than $f<0.5$. After this, Tri/Psc is the best candidate with $\sim 0.9$ gaps expected at this depth. Next, Pal 5 and GD-1 have a similar number of gaps with $~0.3$ expected. Finally, Phoenix and ATLAS both appear to be poor candidates for detecting subhaloes with $0.02$ and $0.01$ gaps at this depth. We note that these predictions rely on the assumed properties in \Tabref{tab:obs_streams}. As a result, these predictions represent a lower bound on the number of gaps since we have assumed that we have observed the full stream length. If these streams are found to be longer than currently observed, the number of gaps would naively increase quadratically with the length of the stream since the number of flybys is proportional to the age of the stream times its length (i.e. \eqref{eq:num_encounters}) and the age itself is proportional to the length. However, the increased age of the stream would also give the gaps more time to grow, potentially making the scaling even stronger. Thus, deep observations of the streams to determine their full length and characterise their densities are critical.

Lastly, we emphasize that this work has only focused on the basic properties of the gaps, such as depth and width. As discussed in \cite{subhalo_properties,bovy_erkal_sanders}, a flyby will also create wiggles in the track of the stream which can be seen in the debris centroid on the sky, the distance to the stream, and the velocities in the stream. An underdensity alone is not sufficient to show a gap is present since other mechanisms like a variable stripping rate can create density variations in the stream. Thus, searches for gaps should also aim to identify these oscillations which will be crucial for constraining the properties of the perturbing subhalo. 

\subsection{Extension to dwarf galaxy streams}

We note that the results of this work cannot be immediately extended to dwarf galaxy streams like Sagittarius \citep{ibata_et_al_2001} or Orphan \citep{orphan} which are substantially longer than the cold streams discussed here. While the flyby rates are still applicable, we would need to determine which flybys can create appreciable gaps. Since streams from disrupting dwarf galaxies have substantially higher velocity dispersions, a larger velocity kick will be needed to form a gap and these streams will not be sensitive to low mass substructure. We expect that the minimum velocity kick needed will likely scale as $m_{\rm prog}^{1/3}$, where $m_{\rm prog}$ is the progenitor mass, just like the stream width and length (e.g. \eqref{eq:width} and \eqref{eq:stream_length}). Since the fraction of velocity kicks above some threshold drops rapidly as the threshold is increased, e.g. \Figref{fig:gap_size_dist_vcut}, it appears unlikely that dwarf galaxy streams will have more gaps than cold streams. 

\subsection{Comparison with other works} \label{sec:comparison}

Let us compare the results of this work against those already reported in the literature. \cite{yoon_etal_2011} and \cite{carlberg_2012} both made similar assumptions as in this work to compute the number of flybys, as well as their relative velocity distribution. As discussed in \Secref{sec:num_flybys}, our derivation for the number of flybys differs slightly from \cite{yoon_etal_2011} but matches \cite{carlberg_2012}. \cite{carlberg_2012} also made predictions for the number of observable gaps which gives significantly more gaps than our model expects. Part of this difference is due to our decreasing the number of subhaloes due to the presence of the Milky Way disk \citep{donghia_et_al_2010}.  

\cite{ngan2014} consider the distribution of gap sizes in N-body simulations and find that the $\Lambda$CDM spectrum of subhaloes primarily create gaps larger than $1^\circ$ \citep[see Fig. 15 of][]{ngan2014}, qualitatively agreeing with the results of this work. \cite{carlberg_2016} simulate a stream on an orbit similar to GD-1 in the presence of subhaloes and find substantially more gaps than expected here. However, their stream age is 10.67 Gyr, significantly older than what we assume for GD-1, and they also perform the gap search over a stream which is $\sim$ 2 radians long, significantly longer than what is observed for GD-1. We note that if we naively scale up our predictions to the age and length of their GD-1, we would get $\sim 5$ gaps deeper than $f<0.9$, in agreement with Figure 7 of \cite{carlberg_2016}.

\subsection{Baryonic effects} \label{sec:baryonic_effects}

In this work we have characterized the frequency of gaps due to dark matter subhaloes. However, as was recently shown in \cite{amorisco_gmcs}, clumps of baryons such as giant molecular clouds (GMCs) can also create gaps in streams. The mass spectrum of GMCs in the Milky Way \citep[e.g.][]{rosolowsky_2005,rice_et_al_2016} shows that within the solar circle, the mass function is a power-law with an index of $\gamma=-1.6$ (not too different from that of the subhaloes) and there are no GMCs above $10^{7} M_\odot$. Outside the solar circle, the mass spectrum is steeper, with $\gamma = -2.1$, and the upper mass drops to $\sim 10^6 M_\odot$. Thus, streams whose pericenters are outside the solar circle should not be strongly affected by GMCs since, as we have shown, the effect from flybys of objects with masses below $10^6 M_\odot$ will not produce a noticeable gap in the stream. However, for streams which pass within the solar circle (i.e. the Pal 5 stream), the most massive GMCs could impart noticeable density fluctuations in the stream. \cite{amorisco_gmcs} evolved N-body realizations of Pal 5 and GD-1 like streams in the presence of the expected distribution of GMCs and reached a similar conclusion with GMCs producing notable gaps in Pal 5 but very few gaps in GD-1. Their analysis also accounted for whether the streams were on a prograde or retrograde with respect to the GMCs and found that prograde orbits produce more substantial gaps since the smaller relative velocities increases the size of the kick from the GMC. Now we can attempt to estimate the importance of GMCs within the framework developed for subhaloes in this work. 

In order to determine the relative importance of these GMCs in the inner part of the Milky Way, we can compare their number density with that of the subhaloes, taking into account what fraction of the stream's orbit is within the disk. Figure 21 of \cite{rice_et_al_2016} shows that their survey found $~40$ GMCs with masses between $10^6-10^7 M_\odot$ within the solar circle. They estimate their completeness by comparing  the total mass of the GMCs they found to the total molecular hydrogen mass in the Milky Way and find they are 28\% complete. Thus, we can estimate that there are $\sim 140$ GMCs in this mass range within the solar circle. If we further assume that the number density of GMCs is constant within the solar circle and that the GMCs are confined to the region within a scale-height of the Milky Way disk, $\sim 250$ pc, we find an average number density of $2.6$ kpc$^{-3}$. In order to compare the effect of the GMCs against the subhaloes, we must estimate the relative number of interactions for a segment of the stream. If a segment travels a length $l$ through a region where perturbers have a number density of $n$ and affect the stream if they pass within a distance $b$, the number of impacts is proportional to $n l b$. If we only consider the effect of the subhaloes within the same mass range as the GMCs considered here, i.e. $10^6-10^7 M_\odot$, then we can assume that $b$ is the same. Thus, we only need to compare the quantity $nl$. The stream is sensitive to subhaloes for its entire orbit, however it is only sensitive to the GMCs when it passes through the disk. For a Pal 5-like stream, if we take the average radius to be 13 kpc and use $n_{\rm sub} = 1.01\times 10^{-3}$kpc$^{-3}$, we get $n_{\rm sub}l_{\rm sub} = 0.08{\rm kpc}^{-2}$ where $l_{\rm sub}$ is the length of the orbit. For the GMCs, if we assume that the orbit passes straight through the disk we get $n_{\rm GMC}l_{\rm GMC} = 1.3{\rm kpc}^{-2}$ where $l_{\rm GMC}$ is twice the disk scale height. Therefore, even though the stream only spends a fraction of its orbit within the disk, the GMCs will have an order of magnitude more interactions and we must consider the effect of GMCs as suggested by \cite{amorisco_gmcs}. Increasing the mass range of the subhaloes we consider does not change this conclusion since the number of gaps created by subhaloes with masses between $10^6-10^7 M_\odot$ is similar to those created by subhaloes between $10^7-10^8 M_\odot$.

This simple analysis has many caveats. First, the number density of GMCs is not constant within the solar circle. In \cite{roman-duval_et_al_2010} the surface density of molecular gas in the Milky Way is shown to peak around 4 kpc and then drop-off: by 7 kpc, the surface density has dropped by an order of magnitude. Thus the relative importance of GMCs depends sensitively on the stream's pericenter. The pericenter of Pal 5 is between $\sim 6-8$ kpc (see \Tabref{tab:obs_streams}) so if the number density of GMCs were 10 times lower, the effect of GMCs and subhaloes would be comparable. Indeed, \cite{amorisco_gmcs} who assumed a pericenter of 8 kpc found that GMCs should produce 0.5 gaps with $\rho/\rho_0<0.71$ in the observed section of Pal 5, similar to our prediction in \Tabref{tab:num_gaps}. Second, this analysis only considered the number of flybys and not the gaps they created. This will be controlled by the relative velocity of the GMCs as compared to the subhaloes and warrants further study \citep[see][]{amorisco_gmcs}. Third, we did not account for the increased path length of a stream through the disk if it is not on a polar orbit, however Pal 5 is relatively close to polar with an orbital inclination of $\sim 65^\circ$ relative to the disk \citep{stream_width}. Fourth, this analysis does not account for the time evolution of GMCs. If the star-formation was stronger in the past, this could increase the number density of GMCs. Finally, this analysis does not account for the finite lifetime of GMCs which are expected to only survive a few free-fall times \citep[e.g.][]{murray_2011}. For the GMCs we consider here that only corresponds to a $\sim 10-20$ Myr and at the typical speeds within the disk, the GMCs will only move a few kpc before dispersing. This can be compared to the timescales over which the GMC would deliver a substantial kick: the region where the accelerations are the largest is on the order of the scale radius for a direct impact so the timescale where the kick is important is on the order of a Myr. Thus, it appears that the disruption of the GMC can safely be neglected. 

Finally we note that in \cite{subhalo_properties} it was shown that given measurements of the density profile of the stream, and two other observables such as the stream track on the sky and the radial velocity along the stream, it is possible to recover the mass and scale radius of the perturber, as well as the time since impact. Thus, in principle, it should be possible to distinguish an impact from a globular cluster from that of a subhalo by the gap properties and by the time since impact. If the interaction occurred within the disk plane, this will lend additional credence to a GMC while if the impact can be convincingly be shown to have occurred far from the disk plane, a subhalo impact will be preferred. 

\subsection{Limitations of the method} \label{sec:limitations}

The method used in this work is based on the perturbation of streams on circular orbits \citep{three_phases}. In this approximation, the stream is treated as being arbitrarily thin and having no velocity dispersion. This neglects the energy and angular momentum dispersion in the stream which can cause the gap depth to plateau as described in \cite{sanders_bovy_erkal_2015}. \cite{sanders_bovy_erkal_2015} also found that the evolution of the gap size depends on where along the stream the impact occurs, with flybys far from the progenitor giving rise to more rapidly growing gaps due to the stretching of the stream itself. Thus, the analysis in this work may be slightly overestimating the depth of the gaps and understimating their size. We have attempted to test this in \Figref{fig:1e6flyby} where we compared the flyby of a $10^6 M_\odot$ in this formalism with the flyby in an effective N-body simulation. This showed a fairly good match indicating that our method is robust. However, if anything our method will over-estimate the depth of gaps and so the number of gaps should be even less than reported in \Tabref{tab:num_gaps}. 

In addition, our method does not account for the change in gap size along the orbit. We have treated the streams as being on circular orbits but naturally a large fraction of them are on orbits with substantial eccentricity. This eccentricity causes the gap size and depth to oscillate as seen in Figures 4,5 of \cite{three_phases} and Figure 13 of \cite{sanders_bovy_erkal_2015}. If we neglect the growth of the gap during an orbit, conservation of angular momentum tells us that the gap sizes goes as $r^{-2}$, e.g. Fig. 13 of \cite{sanders_bovy_erkal_2015}. The gap depth relative to the unperturbed stream exhibits a weaker oscillation but the gaps are deepest at pericenter and shallowest at apocenter. Thus, the predictions in \Tabref{tab:num_gaps} should be seen as an average of the number of gaps expected. At pericenter these gaps will be easier to detect and at apocenter they will be more difficult to spot. We note that both of these limitations are addressed in \cite{bovy_erkal_sanders} where they find broadly similar conclusions. 

Finally, this method does not account for the ongoing disruption of subhaloes. In the regions of the potential where a globular cluster can be tidally stripped, the subhaloes should be disrupting much more vigorously due to their lower density, resulting in dark matter streams as discussed in \cite{bovy_2016}. Including the effect of these partially disrupted subhaloes will create shallower gaps, further lowering the expected number of gaps. 

\section{Conclusion} \label{sec:conclusion}

In this work we have made a prediction for the expected number of stream gaps created by subhaloes and found far fewer gaps than previously expected. This prediction is based on counting the number of subhalo flybys near the stream, similar to the approaches of \cite{yoon_etal_2011} and \cite{carlberg_2012}, and a model for the growth of the resulting gap created by each flyby from \cite{three_phases}. The model for the rate and properties of the flybys in \Secref{sec:flyby_properties} is broadly similar to that in \cite{yoon_etal_2011} and \cite{carlberg_2012} but we expect significantly fewer flybys with a hotter relative velocity distribution. This is partially due to an updated derivation, and partially due to accounting for the depletion of subhaloes by the Milky Way disk \citep[e.g.][]{donghia_et_al_2010}. 

The rate and properties of the flybys are then combined with the analytic model for gap growth described in \Secref{sec:method}. While this analytic prescription is based on perturbations of streams on circular orbits and neglects the dispersion in a real stream, the tests performed in \Secref{sec:method} indicate that it is relatively robust. Using the Pal 5 stream as an example, the distribution of gap sizes and depths is examined in \Secref{sec:gap_properties} and we find several interesting results. First, the gap sizes are larger than previously expected with the majority of the gaps in Pal 5 having typical size of $\sim 5^\circ$ and many as large as $\sim 10^\circ$ and above (see \figref{fig:gap_size_dist_LCDM}). As a result, any searches for gaps in Pal 5 should focus on sizes larger than $\sim 1^\circ$. Second, for a given age of the stream, each perturber mass gives rise to a characteristic gap size (see \figref{fig:gap_size_dist_vary_M}). This can be used to roughly estimate a perturber mass from the size of a gap. Third, we find that the typical gap size is larger for older streams since these gaps have had more time to grow (see \figref{fig:gap_size_dist_vary_time}). 

This formalism was also used to make predictions for the number of gaps in six cold streams around the Milky Way. These predictions are summarised in \Tabref{tab:num_gaps} where it is clear that most streams will have very few gaps. Pal 5 is expected to have 0.3 and 0.7 gaps deeper than $f<0.5$ and $f<0.75$ respectively. As a result, the null detection reported in \cite{ibata_et_al_pal5} is not surprising. This should be contrasted with the 6 gaps detected in \cite{carlberg_pal5_2012}. Indeed, \cite{thomas_et_al_2016} argue that the claimed detections are due to a combination of variation in the Milky Way background with a smooth stream density. The six streams are ranked by the expected number of gaps and the Tri/Psc stream appears to be the most promising candidate with 0.9 and 1.6 gaps deeper than $f<0.5$ and $f<0.75$ respectively. GD-1 is also a promising candidate with a similar number of gaps to Pal 5. Substantially fewer gaps are expected in the ATLAS and Phoenix stream due to their short length and young age. Finally, if the Styx stream is a cold stream with the reported length then it would have the most gaps. 

In addition to the total number of gaps, we also investigate the contribution from each mass decade of subhaloes for Pal 5 and Tri/Psc in \Figref{fig:expected_gaps_Pal5,fig:expected_gaps_tripsc} respectively. This shows that the vast majority of gaps are due to subhaloes with masses in the range $10^6 M_\odot < M < 10^8 M_\odot$. This is a previously unexplored mass range and the detection of even a single confirmed subhalo in this range would be an important test of $\Lambda$CDM and would improve constraints on the mass of a warm dark matter particle. In \Secref{sec:baryonic_effects} we estimate the number of gaps created by GMCs and find that for streams which enter the solar circle, they will be comparable to the number from subhaloes, in agreement with \cite{amorisco_gmcs}. 

While these predictions may appear to dampen the prospects of using cold streams to detect subhaloes, they should instead be thought of as setting realistic expectations for the number of gaps and their properties. With exquisite observations of streams now possible as demonstrated in \cite{ibata_et_al_pal5}, these predictions show that a lack of gaps in Pal 5 is unsurprising but also imply that the search for gaps should be performed on larger scales. Our results suggest that in the near future, deep observations of GD-1 and Tri/Psc, combined with the existing observations of Pal 5, should allow us to begin to uncover the presence of dark subhaloes expected in $\Lambda$CDM. 

\section*{Acknowledgements}

We thank the anonymous referee for their helpful comments. We thank the Streams club at Cambridge for valuable discussions and in particular Sergey Koposov and Thomas de Boer. The research leading to these results has received funding from the European Research Council under the European Union's Seventh Framework Programme (FP/2007-2013)/ERC Grant Agreement no. 308024. J.B. received financial support from the Natural Sciences and Engineering Research Council of Canada. JLS acknowledges the support of the Science and Technology Facilities Council (STFC).

\appendix

\section{Number of subhaloes entering cylinder} \label{sec:appendix}

It is also possible to come up with a simple expression for the number of subhaloes entering the caps of the stream. From the end caps on the left and right side of the cylinder in \Figref{fig:cylinder}, we would expect
\eq{ dN^{\rm L,R}_{\rm enc} = \pi b_{\rm max}^2 \times (|v_s-v_z|dt) \times n_{\rm sub} \times  P(v_z) dv_z ,}
subhaloes to enter the region within $b$ of the stream in time $dt$. Both of these must be integrated over the subhaloes which enter the stream, i.e. on the left side we consider $v_z < v_s$ and on the right side we consider $v_z > v_s$. Performing these integrals over $v_z$ for both endcaps and summing the result, we get
\eq{ \frac{d N^L_{\rm enc}}{dt}+\frac{d N^R_{\rm enc}}{dt} = \pi b_{\rm max}^2 n \sigma \left( \sqrt{\frac{2}{\pi}} \exp(-\frac{v_s^2}{2\sigma^2})  + \frac{v_s  }{\sigma} {\rm erf}(\frac{v_s}{\sigma \sqrt{2}}) \right) \label{eq:end_caps}}

\bibliographystyle{mn2e_long}
\bibliography{citations_pal5}

\begin{thebibliography}{71}
\expandafter\ifx\csname natexlab\endcsname\relax\def\natexlab#1{#1}\fi

\bibitem[{{Abazajian} {et~al}\mbox{.}(2004){Abazajian}, {Adelman-McCarthy},
  {Ag{\"u}eros}, {Allam}, {Anderson}, {Anderson}, {Annis}, {Bahcall}, {Baldry},
  {Bastian}, {Berlind}, {Bernardi}, {Blanton}, {Bochanski}, {Boroski},
  {Briggs}, {Brinkmann}, {Brunner}, {Budav{\'a}ri}, {Carey}, {Carliles},
  {Castander}, {Connolly}, {Csabai}, {Doi}, {Dong}, {Eisenstein}, {Evans},
  {Fan}, {Finkbeiner}, {Friedman}, {Frieman}, {Fukugita}, {Gal}, {Gillespie},
  {Glazebrook}, {Gray}, {Grebel}, {Gunn}, {Gurbani}, {Hall}, {Hamabe},
  {Harris}, {Harris}, {Harvanek}, {Heckman}, {Hendry}, {Hennessy}, {Hindsley},
  {Hogan}, {Hogg}, {Holmgren}, {Ichikawa}, {Ichikawa}, {Ivezi{\'c}}, {Jester},
  {Johnston}, {Jorgensen}, {Kent}, {Kleinman}, {Knapp}, {Kniazev}, {Kron},
  {Krzesinski}, {Kunszt}, {Kuropatkin}, {Lamb}, {Lampeitl}, {Lee}, {Leger},
  {Li}, {Lin}, {Loh}, {Long}, {Loveday}, {Lupton}, {Malik}, {Margon},
  {Matsubara}, {McGehee}, {McKay}, {Meiksin}, {Munn}, {Nakajima}, {Nash},
  {Neilsen}, {Newberg}, {Newman}, {Nichol}, {Nicinski}, {Nieto-Santisteban},
  {Nitta}, {Okamura}, {O'Mullane}, {Ostriker}, {Owen}, {Padmanabhan},
  {Peoples}, {Pier}, {Pope}, {Quinn}, {Richards}, {Richmond}, {Rix}, {Rockosi},
  {Schlegel}, {Schneider}, {Scranton}, {Sekiguchi}, {Seljak}, {Sergey},
  {Sesar}, {Sheldon}, {Shimasaku}, {Siegmund}, {Silvestri}, {Smith}, {Smol{\v
  c}i{\'c}}, {Snedden}, {Stebbins}, {Stoughton}, {Strauss}, {SubbaRao},
  {Szalay}, {Szapudi}, {Szkody}, {Szokoly}, {Tegmark}, {Teodoro}, {Thakar},
  {Tremonti}, {Tucker}, {Uomoto}, {Vanden Berk}, {Vandenberg}, {Vogeley},
  {Voges}, {Vogt}, {Walkowicz}, {Wang}, {Weinberg}, {West}, {White}, {Wilhite},
  {Xu}, {Yanny}, {Yasuda}, {Yip}, {Yocum}, {York}, {Zehavi}, {Zibetti}, \&
  {Zucker}}]{sdss_dr2}
{Abazajian} K. {et~al.}, 2004, \aj, 128, 502

\bibitem[{{Ahn} {et~al}\mbox{.}(2012){Ahn}, {Alexandroff}, {Allende Prieto},
  {Anderson}, {Anderton}, {Andrews}, {Aubourg}, {Bailey}, {Balbinot}, {Barnes},
  \& et~al.}]{sdssdr9}
{Ahn} C.~P. {et~al.}, 2012, ApJS, 203, 21

\bibitem[{{Amorisco}(2015)}]{amorisco_2014}
{Amorisco} N.~C., 2015, \mnras, 450, 575

\bibitem[{{Amorisco} {et~al}\mbox{.}(2016){Amorisco}, {G{\`o}mez}, {Vegetti},
  \& {White}}]{amorisco_gmcs}
{Amorisco} N.~C., {G{\`o}mez} F.~A., {Vegetti} S., {White} S.~D.~M., 2016,
  arXiv:1606.02715

\bibitem[{{Annis} {et~al}\mbox{.}(2014){Annis}, {Soares-Santos}, {Strauss},
  {Becker}, {Dodelson}, {Fan}, {Gunn}, {Hao}, {Ivezi{\'c}}, {Jester}, {Jiang},
  {Johnston}, {Kubo}, {Lampeitl}, {Lin}, {Lupton}, {Miknaitis}, {Seo}, {Simet},
  \& {Yanny}}]{annis2014}
{Annis} J. {et~al.}, 2014, \apj, 794, 120

\bibitem[{{Balbinot} {et~al}\mbox{.}(2016){Balbinot}, {Yanny}, {Li},
  {Santiago}, {Marshall}, {Finley}, {Pieres}, {Abbott}, {Abdalla}, {Allam},
  {Benoit-L{\'e}vy}, {Bernstein}, {Bertin}, {Brooks}, {Burke}, {Carnero
  Rosell}, {Carrasco Kind}, {Carretero}, {Cunha}, {da Costa}, {DePoy}, {Desai},
  {Diehl}, {Doel}, {Estrada}, {Flaugher}, {Frieman}, {Gerdes}, {Gruen},
  {Gruendl}, {Honscheid}, {James}, {Kuehn}, {Kuropatkin}, {Lahav}, {March},
  {Martini}, {Miquel}, {Nichol}, {Ogando}, {Romer}, {Sanchez}, {Schubnell},
  {Sevilla-Noarbe}, {Smith}, {Soares-Santos}, {Sobreira}, {Suchyta}, {Tarle},
  {Thomas}, {Tucker}, {Walker}, \& {The DES Collaboration}}]{phoenix_disc}
{Balbinot} E. {et~al.}, 2016, \apj, 820, 58

\bibitem[{{Battaglia} {et~al}\mbox{.}(2005){Battaglia}, {Helmi}, {Morrison},
  {Harding}, {Olszewski}, {Mateo}, {Freeman}, {Norris}, \&
  {Shectman}}]{battaglia_et_al_2005}
{Battaglia} G. {et~al.}, 2005, \mnras, 364, 433

\bibitem[{{Belokurov}(2013)}]{belokurov2013}
{Belokurov} V., 2013, NARev, 57, 100

\bibitem[{{Belokurov} {et~al}\mbox{.}(2007){Belokurov}, {Evans}, {Irwin},
  {Lynden-Bell}, {Yanny}, {Vidrih}, {Gilmore}, {Seabroke}, {Zucker},
  {Wilkinson}, {Hewett}, {Bramich}, {Fellhauer}, {Newberg}, {Wyse}, {Beers},
  {Bell}, {Barentine}, {Brinkmann}, {Cole}, {Pan}, \& {York}}]{orphan}
{Belokurov} V. {et~al.}, 2007, \apj, 658, 337

\bibitem[{{Bonaca}, {Geha} \& {Kallivayalil}(2012){Bonaca}, {Geha}, \&
  {Kallivayalil}}]{bonacadisc}
{Bonaca} A., {Geha} M., {Kallivayalil} N., 2012, \apjl, 760, L6

\bibitem[{{Bovy}(2014)}]{bovy_2014}
{Bovy} J., 2014, \apj, 795, 95

\bibitem[{{Bovy}(2015)}]{bovy_galpy}
{Bovy} J., 2015, \apjs, 216, 29

\bibitem[{{Bovy}(2016)}]{bovy_2016}
{Bovy} J., 2016, Physical Review Letters, 116, 121301

\bibitem[{{Bovy} {et~al}\mbox{.}(2012){Bovy}, {Allende Prieto}, {Beers},
  {Bizyaev}, {da Costa}, {Cunha}, {Ebelke}, {Eisenstein}, {Frinchaboy},
  {Garc{\'{\i}}a P{\'e}rez}, {Girardi}, {Hearty}, {Hogg}, {Holtzman}, {Maia},
  {Majewski}, {Malanushenko}, {Malanushenko}, {M{\'e}sz{\'a}ros}, {Nidever},
  {O'Connell}, {O'Donnell}, {Oravetz}, {Pan}, {Rocha-Pinto}, {Schiavon},
  {Schneider}, {Schultheis}, {Skrutskie}, {Smith}, {Weinberg}, {Wilson}, \&
  {Zasowski}}]{bovy_et_al_2012}
{Bovy} J. {et~al.}, 2012, \apj, 759, 131

\bibitem[{{Bovy}, {Erkal} \& {Sanders}(2016){Bovy}, {Erkal}, \&
  {Sanders}}]{bovy_erkal_sanders}
{Bovy} J., {Erkal} D., {Sanders} J.~L., 2016, arXiv:1606.03470

\bibitem[{{Bowden}, {Belokurov} \& {Evans}(2015){Bowden}, {Belokurov}, \&
  {Evans}}]{bowden_et_al_gd1}
{Bowden} A., {Belokurov} V., {Evans} N.~W., 2015, \mnras, 449, 1391

\bibitem[{{Carlberg}(2009)}]{carlberg_2009}
{Carlberg} R.~G., 2009, \apjl, 705, L223

\bibitem[{{Carlberg}(2012)}]{carlberg_2012}
{Carlberg} R.~G., 2012, \apj, 748, 20

\bibitem[{{Carlberg}(2016)}]{carlberg_2016}
{Carlberg} R.~G., 2016, \apj, 820, 45

\bibitem[{{Carlberg} \& {Grillmair}(2013)}]{carlberg_gd1_2013}
{Carlberg} R.~G., {Grillmair} C.~J., 2013, \apj, 768, 171

\bibitem[{{Carlberg}, {Grillmair} \& {Hetherington}(2012){Carlberg},
  {Grillmair}, \& {Hetherington}}]{carlberg_pal5_2012}
{Carlberg} R.~G., {Grillmair} C.~J., {Hetherington} N., 2012, \apj, 760, 75

\bibitem[{{Dalal} \& {Kochanek}(2002)}]{Dalal2002}
{Dalal} N., {Kochanek} C.~S., 2002, \apj, 572, 25

\bibitem[{{Deason} {et~al}\mbox{.}(2012){Deason}, {Belokurov}, {Evans}, \&
  {An}}]{deason_et_al_2012}
{Deason} A.~J., {Belokurov} V., {Evans} N.~W., {An} J., 2012, \mnras, 424, L44

\bibitem[{{Diemand}, {Kuhlen} \& {Madau}(2007){Diemand}, {Kuhlen}, \&
  {Madau}}]{diemand_et_al_2007}
{Diemand} J., {Kuhlen} M., {Madau} P., 2007, \apj, 667, 859

\bibitem[{{Diemand} {et~al}\mbox{.}(2008){Diemand}, {Kuhlen}, {Madau}, {Zemp},
  {Moore}, {Potter}, \& {Stadel}}]{Diemand2008}
{Diemand} J., {Kuhlen} M., {Madau} P., {Zemp} M., {Moore} B., {Potter} D.,
  {Stadel} J., 2008, \nat, 454, 735

\bibitem[{{Diemand}, {Moore} \& {Stadel}(2004){Diemand}, {Moore}, \&
  {Stadel}}]{diemand_et_al_2004}
{Diemand} J., {Moore} B., {Stadel} J., 2004, \mnras, 352, 535

\bibitem[{{D'Onghia} {et~al}\mbox{.}(2010){D'Onghia}, {Springel}, {Hernquist},
  \& {Keres}}]{donghia_et_al_2010}
{D'Onghia} E., {Springel} V., {Hernquist} L., {Keres} D., 2010, \apj, 709, 1138

\bibitem[{{Erkal} \& {Belokurov}(2015{\natexlab{a}})}]{three_phases}
{Erkal} D., {Belokurov} V., 2015{\natexlab{a}}, \mnras, 450, 1136

\bibitem[{{Erkal} \& {Belokurov}(2015{\natexlab{b}})}]{subhalo_properties}
{Erkal} D., {Belokurov} V., 2015{\natexlab{b}}, \mnras, 454, 3542

\bibitem[{{Erkal}, {Sanders} \& {Belokurov}(2016){Erkal}, {Sanders}, \&
  {Belokurov}}]{stream_width}
{Erkal} D., {Sanders} J.~L., {Belokurov} V., 2016, \mnras, 461, 1590

\bibitem[{{Fardal}, {Huang} \& {Weinberg}(2015){Fardal}, {Huang}, \&
  {Weinberg}}]{fardal2015}
{Fardal} M.~A., {Huang} S., {Weinberg} M.~D., 2015, \mnras, 452, 301

\bibitem[{{Gibbons}, {Belokurov} \& {Evans}(2014){Gibbons}, {Belokurov}, \&
  {Evans}}]{gibbons_et_al_2014}
{Gibbons} S.~L.~J., {Belokurov} V., {Evans} N.~W., 2014, \mnras, 445, 3788

\bibitem[{{Grillmair}(2009)}]{g_acheron_styx_cocytos_lethe}
{Grillmair} C.~J., 2009, \apj, 693, 1118

\bibitem[{{Grillmair}(2014)}]{grillmair_hermus_hyllus}
{Grillmair} C.~J., 2014, \apjl, 790, L10

\bibitem[{{Grillmair} \& {Carlberg}(2016)}]{grillmair_carlberg_2016}
{Grillmair} C.~J., {Carlberg} R.~G., 2016, \apjl, 820, L27

\bibitem[{{Grillmair} \& {Carlin}(2016)}]{stream_book_grillmair_carlin}
{Grillmair} C.~J., {Carlin} J.~L., 2016, in Astrophysics and Space Science
  Library, Vol. 420, Astrophysics and Space Science Library, {Newberg} H.~J.,
  {Carlin} J.~L., eds., p.~87

\bibitem[{{Grillmair} \& {Dionatos}(2006{\natexlab{a}})}]{gd_pal5}
{Grillmair} C.~J., {Dionatos} O., 2006{\natexlab{a}}, \apjl, 641, L37

\bibitem[{{Grillmair} \& {Dionatos}(2006{\natexlab{b}})}]{gd1disc}
{Grillmair} C.~J., {Dionatos} O., 2006{\natexlab{b}}, \apjl, 643, L17

\bibitem[{{Hezaveh} {et~al}\mbox{.}(2013){Hezaveh}, {Dalal}, {Holder},
  {Kuhlen}, {Marrone}, {Murray}, \& {Vieira}}]{Hezaveh2013}
{Hezaveh} Y., {Dalal} N., {Holder} G., {Kuhlen} M., {Marrone} D., {Murray} N.,
  {Vieira} J., 2013, \apj, 767, 9

\bibitem[{{Hezaveh} {et~al}\mbox{.}(2016){Hezaveh}, {Dalal}, {Marrone}, {Mao},
  {Morningstar}, {Wen}, {Blandford}, {Carlstrom}, {Fassnacht}, {Holder},
  {Kemball}, {Marshall}, {Murray}, {Perreault Levasseur}, {Vieira}, \&
  {Wechsler}}]{Hezaveh2016}
{Hezaveh} Y.~D. {et~al.}, 2016, \apj, 823, 37

\bibitem[{{Ibata} {et~al}\mbox{.}(2001){Ibata}, {Lewis}, {Irwin}, {Totten}, \&
  {Quinn}}]{ibata_et_al_2001}
{Ibata} R., {Lewis} G.~F., {Irwin} M., {Totten} E., {Quinn} T., 2001, \apj,
  551, 294

\bibitem[{{Ibata} {et~al}\mbox{.}(2002){Ibata}, {Lewis}, {Irwin}, \&
  {Quinn}}]{ibata_et_al_2002}
{Ibata} R.~A., {Lewis} G.~F., {Irwin} M.~J., {Quinn} T., 2002, \mnras, 332, 915

\bibitem[{{Ibata}, {Lewis} \& {Martin}(2016){Ibata}, {Lewis}, \&
  {Martin}}]{ibata_et_al_pal5}
{Ibata} R.~A., {Lewis} G.~F., {Martin} N.~F., 2016, \apj, 819, 1

\bibitem[{{Johnston}, {Spergel} \& {Haydn}(2002){Johnston}, {Spergel}, \&
  {Haydn}}]{johnston_et_al_2002}
{Johnston} K.~V., {Spergel} D.~N., {Haydn} C., 2002, \apj, 570, 656

\bibitem[{{Koposov} {et~al}\mbox{.}(2008){Koposov}, {Belokurov}, {Evans},
  {Hewett}, {Irwin}, {Gilmore}, {Zucker}, {Rix}, {Fellhauer}, {Bell}, \&
  {Glushkova}}]{koposov2008}
{Koposov} S. {et~al.}, 2008, \apj, 686, 279

\bibitem[{{Koposov} {et~al}\mbox{.}(2012){Koposov}, {Belokurov}, {Evans},
  {Gilmore}, {Gieles}, {Irwin}, {Lewis}, {Niederste-Ostholt}, {Pe{\~n}arrubia},
  {Smith}, {Bizyaev}, {Malanushenko}, {Malanushenko}, {Schneider}, \&
  {Wyse}}]{koposov_2012}
{Koposov} S.~E. {et~al.}, 2012, \apj, 750, 80

\bibitem[{{Koposov} {et~al}\mbox{.}(2014){Koposov}, {Irwin}, {Belokurov},
  {Gonzalez-Solares}, {Yoldas}, {Lewis}, {Metcalfe}, \& {Shanks}}]{atlasdisc}
{Koposov} S.~E., {Irwin} M., {Belokurov} V., {Gonzalez-Solares} E., {Yoldas}
  A.~K., {Lewis} J., {Metcalfe} N., {Shanks} T., 2014, \mnras, 442, L85

\bibitem[{{K{\"u}pper} {et~al}\mbox{.}(2015){K{\"u}pper}, {Balbinot}, {Bonaca},
  {Johnston}, {Hogg}, {Kroupa}, \& {Santiago}}]{kuepper_et_al_pal5}
{K{\"u}pper} A.~H.~W., {Balbinot} E., {Bonaca} A., {Johnston} K.~V., {Hogg}
  D.~W., {Kroupa} P., {Santiago} B.~X., 2015, \apj, 803, 80

\bibitem[{{Mao} \& {Schneider}(1998)}]{Mao1998}
{Mao} S., {Schneider} P., 1998, \mnras, 295, 587

\bibitem[{{Martin} {et~al}\mbox{.}(2013){Martin}, {Carlin}, {Newberg}, \&
  {Grillmair}}]{tripsc_charlesmartin}
{Martin} C., {Carlin} J.~L., {Newberg} H.~J., {Grillmair} C., 2013, \apjl, 765,
  L39

\bibitem[{{Martin} {et~al}\mbox{.}(2014){Martin}, {Ibata}, {Rich}, {Collins},
  {Fardal}, {Irwin}, {Lewis}, {McConnachie}, {Babul}, {Bate}, {Chapman},
  {Conn}, {Crnojevi{\'c}}, {Ferguson}, {Mackey}, {Navarro}, {Pe{\~n}arrubia},
  {Tanvir}, \& {Valls-Gabaud}}]{tripsc_pandas}
{Martin} N.~F. {et~al.}, 2014, \apj, 787, 19

\bibitem[{{Murray}(2011)}]{murray_2011}
{Murray} N., 2011, \apj, 729, 133

\bibitem[{{Ngan} \& {Carlberg}(2014)}]{ngan2014}
{Ngan} W.~H.~W., {Carlberg} R.~G., 2014, \apj, 788, 181

\bibitem[{{Odenkirchen} {et~al}\mbox{.}(2003){Odenkirchen}, {Grebel}, {Dehnen},
  {Rix}, {Yanny}, {Newberg}, {Rockosi}, {Mart{\'{\i}}nez-Delgado}, {Brinkmann},
  \& {Pier}}]{pal5disc}
{Odenkirchen} M. {et~al.}, 2003, \aj, 126, 2385

\bibitem[{{Padmanabhan} {et~al}\mbox{.}(2008){Padmanabhan}, {Schlegel},
  {Finkbeiner}, {Barentine}, {Blanton}, {Brewington}, {Gunn}, {Harvanek},
  {Hogg}, {Ivezi{\'c}}, {Johnston}, {Kent}, {Kleinman}, {Knapp}, {Krzesinski},
  {Long}, {Neilsen}, {Nitta}, {Loomis}, {Lupton}, {Roweis}, {Snedden},
  {Strauss}, \& {Tucker}}]{sdss_ubercal}
{Padmanabhan} N. {et~al.}, 2008, \apj, 674, 1217

\bibitem[{{Piffl}, {Penoyre} \& {Binney}(2015){Piffl}, {Penoyre}, \&
  {Binney}}]{piffl_et_al_2015}
{Piffl} T., {Penoyre} Z., {Binney} J., 2015, \mnras, 451, 639

\bibitem[{{Rice} {et~al}\mbox{.}(2016){Rice}, {Goodman}, {Bergin}, {Beaumont},
  \& {Dame}}]{rice_et_al_2016}
{Rice} T.~S., {Goodman} A.~A., {Bergin} E.~A., {Beaumont} C., {Dame} T.~M.,
  2016, \apj, 822, 52

\bibitem[{{Roman-Duval} {et~al}\mbox{.}(2010){Roman-Duval}, {Jackson}, {Heyer},
  {Rathborne}, \& {Simon}}]{roman-duval_et_al_2010}
{Roman-Duval} J., {Jackson} J.~M., {Heyer} M., {Rathborne} J., {Simon} R.,
  2010, \apj, 723, 492

\bibitem[{{Rosolowsky}(2005)}]{rosolowsky_2005}
{Rosolowsky} E., 2005, \pasp, 117, 1403

\bibitem[{{Sanders}(2014)}]{sanders2014}
{Sanders} J.~L., 2014, \mnras, 443, 423

\bibitem[{{Sanders}, {Bovy} \& {Erkal}(2016){Sanders}, {Bovy}, \&
  {Erkal}}]{sanders_bovy_erkal_2015}
{Sanders} J.~L., {Bovy} J., {Erkal} D., 2016, \mnras, 457, 3817

\bibitem[{{Shanks} {et~al}\mbox{.}(2015){Shanks}, {Metcalfe}, {Chehade},
  {Findlay}, {Irwin}, {Gonzalez-Solares}, {Lewis}, {Yoldas}, {Mann}, {Read},
  {Sutorius}, \& {Voutsinas}}]{shanks2015}
{Shanks} T. {et~al.}, 2015, \mnras, 451, 4238

\bibitem[{{Siegal-Gaskins} \& {Valluri}(2008)}]{siegal_valluri_2008}
{Siegal-Gaskins} J.~M., {Valluri} M., 2008, \apj, 681, 40

\bibitem[{{Springel}(2005)}]{springel_2005}
{Springel} V., 2005, \mnras, 364, 1105

\bibitem[{{Springel} {et~al}\mbox{.}(2008){Springel}, {Wang}, {Vogelsberger},
  {Ludlow}, {Jenkins}, {Helmi}, {Navarro}, {Frenk}, \&
  {White}}]{springel_et_al_2008}
{Springel} V. {et~al.}, 2008, \mnras, 391, 1685

\bibitem[{{Stoughton} {et~al}\mbox{.}(2002){Stoughton}, {Lupton}, {Bernardi},
  {Blanton}, {Burles}, {Castander}, {Connolly}, {Eisenstein}, {Frieman},
  {Hennessy}, {Hindsley}, {Ivezi{\'c}}, {Kent}, {Kunszt}, {Lee}, {Meiksin},
  {Munn}, {Newberg}, {Nichol}, {Nicinski}, {Pier}, {Richards}, {Richmond},
  {Schlegel}, {Smith}, {Strauss}, {SubbaRao}, {Szalay}, {Thakar}, {Tucker},
  {Vanden Berk}, {Yanny}, {Adelman}, {Anderson}, {Anderson}, {Annis},
  {Bahcall}, {Bakken}, {Bartelmann}, {Bastian}, {Bauer}, {Berman},
  {B{\"o}hringer}, {Boroski}, {Bracker}, {Briegel}, {Briggs}, {Brinkmann},
  {Brunner}, {Carey}, {Carr}, {Chen}, {Christian}, {Colestock}, {Crocker},
  {Csabai}, {Czarapata}, {Dalcanton}, {Davidsen}, {Davis}, {Dehnen},
  {Dodelson}, {Doi}, {Dombeck}, {Donahue}, {Ellman}, {Elms}, {Evans}, {Eyer},
  {Fan}, {Federwitz}, {Friedman}, {Fukugita}, {Gal}, {Gillespie}, {Glazebrook},
  {Gray}, {Grebel}, {Greenawalt}, {Greene}, {Gunn}, {de Haas}, {Haiman},
  {Haldeman}, {Hall}, {Hamabe}, {Hansen}, {Harris}, {Harris}, {Harvanek},
  {Hawley}, {Hayes}, {Heckman}, {Helmi}, {Henden}, {Hogan}, {Hogg}, {Holmgren},
  {Holtzman}, {Huang}, {Hull}, {Ichikawa}, {Ichikawa}, {Johnston}, {Kauffmann},
  {Kim}, {Kimball}, {Kinney}, {Klaene}, {Kleinman}, {Klypin}, {Knapp},
  {Korienek}, {Krolik}, {Kron}, {Krzesi{\'n}ski}, {Lamb}, {Leger},
  {Limmongkol}, {Lindenmeyer}, {Long}, {Loomis}, {Loveday}, {MacKinnon},
  {Mannery}, {Mantsch}, {Margon}, {McGehee}, {McKay}, {McLean}, {Menou},
  {Merelli}, {Mo}, {Monet}, {Nakamura}, {Narayanan}, {Nash}, {Neilsen},
  {Newman}, {Nitta}, {Odenkirchen}, {Okada}, {Okamura}, {Ostriker}, {Owen},
  {Pauls}, {Peoples}, {Peterson}, {Petravick}, {Pope}, {Pordes}, {Postman},
  {Prosapio}, {Quinn}, {Rechenmacher}, {Rivetta}, {Rix}, {Rockosi}, {Rosner},
  {Ruthmansdorfer}, {Sandford}, {Schneider}, {Scranton}, {Sekiguchi}, {Sergey},
  {Sheth}, {Shimasaku}, {Smee}, {Snedden}, {Stebbins}, {Stubbs}, {Szapudi},
  {Szkody}, {Szokoly}, {Tabachnik}, {Tsvetanov}, {Uomoto}, {Vogeley}, {Voges},
  {Waddell}, {Walterbos}, {Wang}, {Watanabe}, {Weinberg}, {White}, {White},
  {Wilhite}, {Wolfe}, {Yasuda}, {York}, {Zehavi}, \& {Zheng}}]{sdss_edr}
{Stoughton} C. {et~al.}, 2002, \aj, 123, 485

\bibitem[{{The Dark Energy Survey Collaboration}(2005)}]{des_survey}
{The Dark Energy Survey Collaboration}, 2005, arXiv:astro-ph/0510346

\bibitem[{{Thomas} {et~al}\mbox{.}(2016){Thomas}, {Ibata}, {Famaey}, {Martin},
  \& {Lewis}}]{thomas_et_al_2016}
{Thomas} G.~F., {Ibata} R., {Famaey} B., {Martin} N.~F., {Lewis} G.~F., 2016,
  arXiv:1605.05520

\bibitem[{{Vegetti} {et~al}\mbox{.}(2010){Vegetti}, {Koopmans}, {Bolton},
  {Treu}, \& {Gavazzi}}]{Vegetti2010}
{Vegetti} S., {Koopmans} L.~V.~E., {Bolton} A., {Treu} T., {Gavazzi} R., 2010,
  \mnras, 408, 1969

\bibitem[{{Vegetti} {et~al}\mbox{.}(2012){Vegetti}, {Lagattuta}, {McKean},
  {Auger}, {Fassnacht}, \& {Koopmans}}]{Vegetti2012}
{Vegetti} S., {Lagattuta} D.~J., {McKean} J.~P., {Auger} M.~W., {Fassnacht}
  C.~D., {Koopmans} L.~V.~E., 2012, \nat, 481, 341

\bibitem[{{Yoon}, {Johnston} \& {Hogg}(2011){Yoon}, {Johnston}, \&
  {Hogg}}]{yoon_etal_2011}
{Yoon} J.~H., {Johnston} K.~V., {Hogg} D.~W., 2011, \apj, 731, 58

\end{thebibliography}

\end{document}